\documentclass[fleqn,usenatbib]{mnras}
\usepackage{etoolbox}
\makeatletter
\patchcmd\@combinedblfloats{\box\@outputbox}{\unvbox\@outputbox}{}{%
   \errmessage{\noexpand\@combinedblfloats could not be patched}%
}%
 \makeatother
\usepackage{newtxtext,newtxmath}
\usepackage[T1]{fontenc}
\usepackage{ae,aecompl}
\usepackage{soul}
\usepackage{graphicx}	% Including figure files
\usepackage{amsmath}	% Advanced maths commands
\usepackage{amssymb}	% Extra maths symbols

\graphicspath{{./}{figures/}}
\newcommand{\hi}{\ifmmode \ion{H}{I} \else $\ion{H}{I}$ \fi}
\newcommand{\hii}{\ifmmode \ion{H}{II} \else $\ion{H}{II}$ \fi}
\newcommand{\hei}{\ifmmode \ion{He}{I} \else $\ion{He}{I}$ \fi}
\newcommand{\heii}{\ifmmode \ion{He}{II} \else $\ion{He}{II}$ \fi}
\newcommand{\heiii}{\ifmmode \ion{He}{III} \else $\ion{He}{III}$ \fi}
\newcommand{\arepo}{{\sc arepo}\ }
\newcommand{\areport}{{\sc arepo-rt}\ }

\title[Temperature Fluctuations in the Ly$\alpha$ Forest]{Imprints of temperature fluctuations on the $z\sim 5$ Lyman-$\alpha$ forest: a view from radiation-hydrodynamic simulations of reionization}

\author[X. H. Wu et al.]{
Xiaohan Wu$^{1}$\thanks{E-mail: xiaohan.wu@cfa.harvard.edu}\href{https://orcid.org/0000-0003-2061-4299}{\includegraphics[scale=0.8]{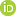}}, 
Matthew McQuinn$^{2}$, 
Rahul Kannan$^{1 \thanks{Einstein Fellow}}$, 
Anson D'Aloisio$^{3}$, 
\newauthor
Simeon Bird$^{3}$,
Federico Marinacci$^{4}$\href{https://orcid.org/0000-0003-3816-7028}{\includegraphics[scale=0.8]{figures/orcid.png}}, 
Romeel Dav{\'e}$^{5,6,7}$, 
Lars Hernquist$^{1}$
\\
\\
$^{1}$Harvard-Smithsonian Center for Astrophysics, 60 Garden Street, Cambridge 02138, MA, USA \\
$^{2}$Astronomy Department, University of Washington, Seattle, WA 98195, USA \\
$^{3}$Department of Physics and Astronomy, University of California, Riverside, CA 92521, USA \\
$^{4}$Department of Physics \& Astronomy, University of Bologna, via Gobetti 93/2, 40129 Bologna, Italy \\
$^{5}$Institute for Astronomy, University of Edinburgh, EH9 3HJ, UK \\
$^{6}$ University of the Western Cape, Bellville, Cape Town 7535, South Africa \\
$^{7}$ South African Astronomical Observatories, Observatory, Cape Town 7925, South Africa
}

\date{Accepted 2019 October 01. Received 2019 September 20; in original form 2019 July 10.}

\pubyear{2019}

\begin{document}
\label{firstpage}
\pagerange{\pageref{firstpage}--\pageref{lastpage}}
\maketitle

\begin{abstract}
Reionization leads to large spatial fluctuations in the intergalactic temperature that can persist well after its completion. We study the imprints of such fluctuations on the $z\sim5$ Ly$\alpha$ forest flux power spectrum using a set of radiation-hydrodynamic simulations that model different reionization scenarios.  We find that large-scale coherent temperature fluctuations bring $\sim20-60\%$ extra power at $k\sim0.002$~s/km, with the largest enhancements in the models where reionization is extended or ends the latest.  On smaller scales ($k\gtrsim0.1$~s/km), we find that temperature fluctuations suppress power by $\lesssim10\%$. We find that the shape of the power spectrum is mostly sensitive to the reionization midpoint rather than temperature fluctuations from reionization's patchiness.  However, for all of our models with reionization midpoints of $z\le 8$ ($z\le 12$) the shape differences are $\lesssim20\%$ ($\lesssim40\%$) because of a surprisingly well-matched cancellation between thermal broadening and pressure smoothing that occurs for realistic thermal histories.  We also consider fluctuations in the ultraviolet background, finding their impact on the power spectrum to be much smaller than temperature fluctuations at $k\gtrsim0.01$~s/km. Furthermore, we compare our models to power spectrum measurements, finding that none of our models with reionization midpoints of $z<8$ is strongly preferred over another and that all of our models with midpoints of $z\geq8$ are excluded at $2.5\sigma$. Future measurements may be able to distinguish between viable reionization models if they can be performed at lower $k$ or, alternatively, if the error bars on the high-$k$ power can be reduced by a factor of $1.5$.
\end{abstract}


%%%%%%%%%%%%%%%%%%%%%%%%%%%%%%%%%%%%%%%%%%%%%%%%%%

%%%%%%%%%%%%%%%%% BODY OF PAPER %%%%%%%%%%%%%%%%%%

\section{Introduction}

The epoch of reionization is the era when radiation from the first sources reionized the intergalactic medium (IGM), turning it from hundreds of degrees Kelvin and neutral to $\sim 20,000$~K and highly ionized \citep[e.g.][]{MER94, McQu12, DAlo18Ifront}. After reionization, the photoheated gas cooled mainly through adiabatic expansion and Compton cooling off of the cosmic microwave background \citep[CMB;][]{MER94, HG97, Hui03, McQu16}.
A patchy reionization where different regions are reionized at different times results in order unity scatter in the IGM temperature field when reionization completes \citep[e.g.][]{Trac08, DAlo15}. 
Temperature fluctuations persist at this level for $\Delta z\sim1-2$ before the IGM settles onto a tight temperature--density relation, which arises from the competition between photoheating of the relic \hi and the cooling processes \citep[e.g.][]{HG97}.

The Ly$\alpha$ forest, caused by Ly$\alpha$ absorption of neutral hydrogen atoms along the line of sight, is a valuable tool for studying the thermal and ionization state of the IGM at $z\lesssim7$.
The Ly$\alpha$ optical depth $\tau$ is proportional to the \hi number density $n_{\rm \hi}$, which scales as $T^{-0.7}\Delta^2/\Gamma$ in photoionized gas. Here $\Delta\equiv\rho/\bar{\rho}$ is the gas density in units of the cosmic mean, $\Gamma$ is the \hi photoionization rate, $T$ is the IGM temperature, and the $T^{-0.7}$ scaling arises from the temperature dependence of the hydrogen recombination rate \citep[e.g.][]{McQu16review}.
Therefore in addition to density fluctuations, temperature fluctuations introduce extra scatter in the Ly$\alpha$ forest opacity. While transmission in the $z<6$ forest segments has been used to place a lower bound on the endpoint of reionization \citep{Fan06, McGr15}, the finer details of how the opacity varies carry information on how reionization proceeded. % For instance, if the large variations in the Ly$\alpha$ opacity observed between $50$Mpc$/h$ regions at $z\sim5-6$ \citep{Fan06, Beck15} is caused by $T$ fluctuations, a late-ending and/or extended reionization is needed to reproduce the observations \citep{DAlo15, Keat18, Kulk19}.
In this work we focus on the imprints of $T$ fluctuations on the $z\sim5$ Ly$\alpha$ forest flux power spectrum.

Temperature fluctuations can alter the shape of correlations in the Ly$\alpha$ forest.
On small scales, thermal broadening erases structures in the forest by widening the absorption lines in velocity space. The smoothing of the gas distribution relative to dark matter by gas pressure also eliminates structures at $\lesssim100$ kpc scales \citep{GH98, Peep10, Kulk15, Nasir16, Onor17sim, Onor17ps}. Structures are also broadened further in velocity space by pressure induced peculiar velocities \citep[e.g.][]{Cen94}. %Because the effect of pressure smoothing is three-dimensional, projection to the line-of-sight flux leaves imprints on larger scales than the purely line-of-sight effect of thermal broadening.
On even larger scales still, the clustering of the ionizing sources and the size of the ionized bubbles during reionization leave their imprints on the Ly$\alpha$ flux via temperature fluctuations. A number of works have shown that large-scale correlations in the IGM temperature field can bring excess power in the lowest observable wavenumbers in the power spectrum \citep{Cen09, DAlo18fluc, Onor18, MC19}, but the small-scale effects of temperature fluctuations have still not been studied in detail. %\citep[but see the analysis in][]{McQu11, Hui17}.

In the standard way of extracting the IGM thermal history or warm/fuzzy dark matter mass from the observed power spectrum, one compares observations to simulations that adopt a uniform UV background \citep[e.g.][]{HM12} with different reionization redshifts \citep[e.g.][]{Viel13, Walt18, Boera18}. If temperature fluctuations affect the small and/or intermediate-scale power significantly, e.g. by increasing the small-scale power \citep{Hui17}, such estimates of the IGM temperature may be biased. Constraints on the warm/fuzzy dark matter mass from the Ly$\alpha$ forest may also be weakened, especially considering that the tightest constraints are derived from $z\gtrsim5$ where these fluctuations are largest \citep[e.g.][]{Viel08, Viel13, Irsic17a, Irsic17b}.
Moreover, if imprints of $T$ fluctuations on the power spectrum (e.g. extra large-scale power) are detectable, they can provide information on when reionization ended, how long it lasted, and the patchiness of this process.
It is thus crucial to understand how $T$ fluctuations change the shape of the power spectrum and the validity of the assumption made by all analyses that the IGM follows a single power law $T-\rho$ relation.

For the first time, we have explored the above mentioned questions by running fully self-consistent radiation-hydrodynamic (RHD) simulations with the state-of-art Illustris galaxy formation model \citep{Voge13, Voge14a, Voge14b, Genel14, Sija15, Nels15}.
Since RHD simulations are able to capture the geometry of the cosmic web and the propagation of ionization fronts (I-front), they can better model the post-reionization IGM temperature. Hydrodynamic simulations coupled with semi-numerical reionization models that assume a single temperature when gas is reionized  \citep[e.g.][]{Onor18} may not capture the strong dependencies of the imparted temperature on the ionization front speed \citep{DAlo18Ifront}.
Moreover, RHD simulations are able to take into account the kinematic response of the gas to the photoheating, which is missed when  post-processing hydrodynamic simulations with radiative transfer \citep[RT; e.g.][]{Keat18}. This is crucial for resolving the pressure smoothing effects, which we find to be as important as thermal broadening for the impact of temperature on the power spectrum.
In addition, our simulations span a larger parameter space of the endpoint and duration of reionization than previous works. With different levels of temperature fluctuations at a given post-reionization redshift, they allow us to understand the distinguishability of different reionization models.

This paper is organized as follows. Sec.~\ref{sec:methods} introduces the simulation setup and the tool for generating mock Ly$\alpha$ forest spectra. Sec.~\ref{sec:results} presents power spectra in different reionization models and how temperature fluctuations leave their imprints. Sec.~\ref{sec:discussions} shows comparisons with other works and discusses whether the current simulations can give constraints on reionization based on recent observational data. Sec.~\ref{sec:conclusions} summarizes this work.

\section{Methods}
\label{sec:methods}
We ran cosmological RHD simulations using the \areport code \citep{Kann19}, an RHD extension of the moving-mesh cosmological hydrodynamic code \arepo \citep{Spri10}. The simulations use the star formation and stellar feedback scheme of the Illustris galaxy formation model \citep{Voge13}, which follows the implementation of \citet{SH03}. In simulations with RT, gas cooling and photoheating are achieved by a non-equilibrium hydrogen and helium thermochemistry network in the RT module, taking into account the radiative cooling processes outlined in \citet{KWH96}. In simulations using a uniform UV background, gas is assumed to be in ionization equilibrium with the background. We refer readers to \citet{Wu19} for a detailed description of the model implementation and briefly summarize the main features concerning this work below.

\areport solves the moment equations of radiative transfer using the M1 closure. We adopt a reduced speed of light \citep{Gned01} of $0.1c$, where $c$ is the actual speed of light. We trace three frequency bins relevant for H and He reionization: $[13.6,18.3]$ eV, $[18.3, 24.6]$ eV, $[24.6, 54.4]$ eV. Appendix~\ref{sec:T_convergence} will show the numerical convergence of the gas temperature with respect to the choice of the reduced speed of light, the number of frequency bins, and the size of gas cells.
Star particles are taken to be the radiation sources in the simulations. Each star particle represents a co-eval, single metallicity stellar population with a \citet{Chab03} IMF. The number of photons emitted by a star particle is calculated via integrating its spectral energy distribution, obtained by interpolating the \citet{BC03} model using the star particle age and metallicity.
In order to take into account unresolved absorption by the birth clouds of the star particles, 
we multiply the luminosities of the star particles by an escape fraction $f^{\rm birth}_{\rm esc}$\footnote{We note that $f^{\rm birth}_{\rm esc}$ is different from the total escape fraction from the galaxy. The latter takes into account absorption due to all gas cells in a galaxy, which is modeled by RT.}. Through adjusting $f^{\rm birth}_{\rm esc}$, we change when reionization ends and how long it lasts.
Specifically, we explore three forms of hydrogen reionization history: early reionization (RT-early, $f^{\rm birth}_{\rm esc} = 1.0$), late reionization (RT-late, $f^{\rm birth}_{\rm esc} = 0.4$), extended reionization (RT-extended, time-varying $f^{\rm birth}_{\rm esc}$). The time-varying $f^{\rm birth}_{\rm esc}$ is given by:
\begin{equation}
  f^{\rm birth}_{\rm esc}(z)=\begin{cases}
               1.0,\ z\ge9\\
               \max(0.3, (z/9)^4),\ z<9
            \end{cases}
  \label{eq:fesc}
\end{equation}
where $f^{\rm birth}_{\rm sec} = 0.3$ occurs at $z\approx6.7$.

The fiducial set of simulations has a box size of $(25\, h^{-1}\ {\rm Mpc})^3$ with $512^3$ dark matter particles and initially $512^3$ gas cells (denoted as L25n512\footnote{Hereafter we use LXXnXX to name our simulations. The numbers following L and n represent the box size and the number of resolution elements, respectively.}). We adopt a \citet{Planck16} cosmology with $\Omega_m = 0.3089$, $\Omega_\Lambda = 0.6911$, $\Omega_b = 0.0486$, $h = 0.6774$, and $\sigma_8 = 0.8159$. The simulations have a dark matter particle mass of $1.2\times10^7\ M_\odot$, and initial gas cell mass of $1.9\times10^6\ M_\odot$. Cells are refined or de-refined in order keep their mass within a factor of 2 from this initial target mass.
All simulations are started from a snapshot at $z=15$. This saves a significant amount of computing time for simulations with RT. Before this redshift we assume that reionization has not started. Indeed, although the first stars form at $z\sim20$, the bulk of reionization occurs below $z\sim12$ in our simulations. The RT simulations are stopped when the volume-averaged \hi fraction falls to $\sim 10^{-3}$. From there we restart the simulations from the last snapshot using the \citet[][hereafter FG09]{FG09} UV background instead of performing RT and run them to $z=5$. This further reduces a significant amount of computational cost.
Adopting a uniform UV background after reionization is a good assumption for our simulations, considering the observed photon mean free path at $z\sim5$ is $\gtrsim40\, h^{-1}$ (comoving) Mpc \citep{Wors14}. The FG09 UV background also allows all simulations to roughly match the observed volume-averaged \hi fraction at $z\sim5-6$ \citep[e.g.][]{Fan06}.
Nevertheless, we present an estimate of the effects of large-scale UV background fluctuations on the power spectrum in Sec.~\ref{sec:UVB_fluc}.  We also note that the thermal evolution after ionization is only weakly sensitive to the details of the post-reionization ionizing background, especially at high redshifts and in low densities as the gas has not relaxed to the thermal asymptote \citep{McQu16}.

In addition to the RT simulations, we perform four L25n512 flash reionization (FR) simulations (named FR-zXX) where we reionize the entire simulation volume at a certain reionization redshift and inject heat to all gas cells, except the star-forming ones, by increasing their temperature by $\Delta T = 20,000$~K. After the heat injection, we turn on the FG09 UV background and let the gas cool and evolve.
Reionization occurs uniformly in these FR simulations such that the gas follows a tight $T-\rho$ relation after reionization.
Two FR simulations have reionization redshifts of $6.7$ and $8.0$, roughly the same as the reionization midpoints of the RT simulations. This is inspired by the results of \citet{Onor18}, which found little difference in the power spectrum at small scales when the midpoint of reionization is fixed.
To explore whether the observed power spectrum favors very high reionization redshifts, we run two FR simulations with reionization redshifts $z=10$ and $z=12$. We use the UV background at $z=8.5$ for the IGM evolution before $z=8.5$ to ensure that the volume-averaged \hi fraction is always $\gtrsim10^{-4}$.
Among these FR simulations, FR-z6.7, FR-z8.0, and FR-z10 are run to $z=4.6$ to examine the comparison of the simulated power spectra with the observations. FR-z12 is stopped at $z=5.0$.

In order to explore box size effects on the simulated power spectrum, we have performed two L37.5n768 RT simulations with the RT-extended and RT-early model variations.
Previous works have shown that a box size of $\gtrsim40\, h^{-1}$ Mpc is preferred in order to suppress the numerical error of the simulated Ly$\alpha$ power spectrum to within $\sim10\%$ \citep{Bolt09, Lukic15}. Larger boxes better capture the clustering of the ionizing sources and the growth of the ionized bubbles, especially at the late stages of reionization. The size of the bubbles roughly sets the scale of correlations in the IGM temperature field, which can be as large as several tens of comoving Mpc near the end of reionization \citep[e.g.][]{McQu07, GK14}. Having the two box sizes allows us to remark on how well the $25\, h^{-1}$ Mpc and $37.5\, h^{-1}$ Mpc simulations statistically sample the structures during reionization.

\begin{figure}
\includegraphics[width=\columnwidth]{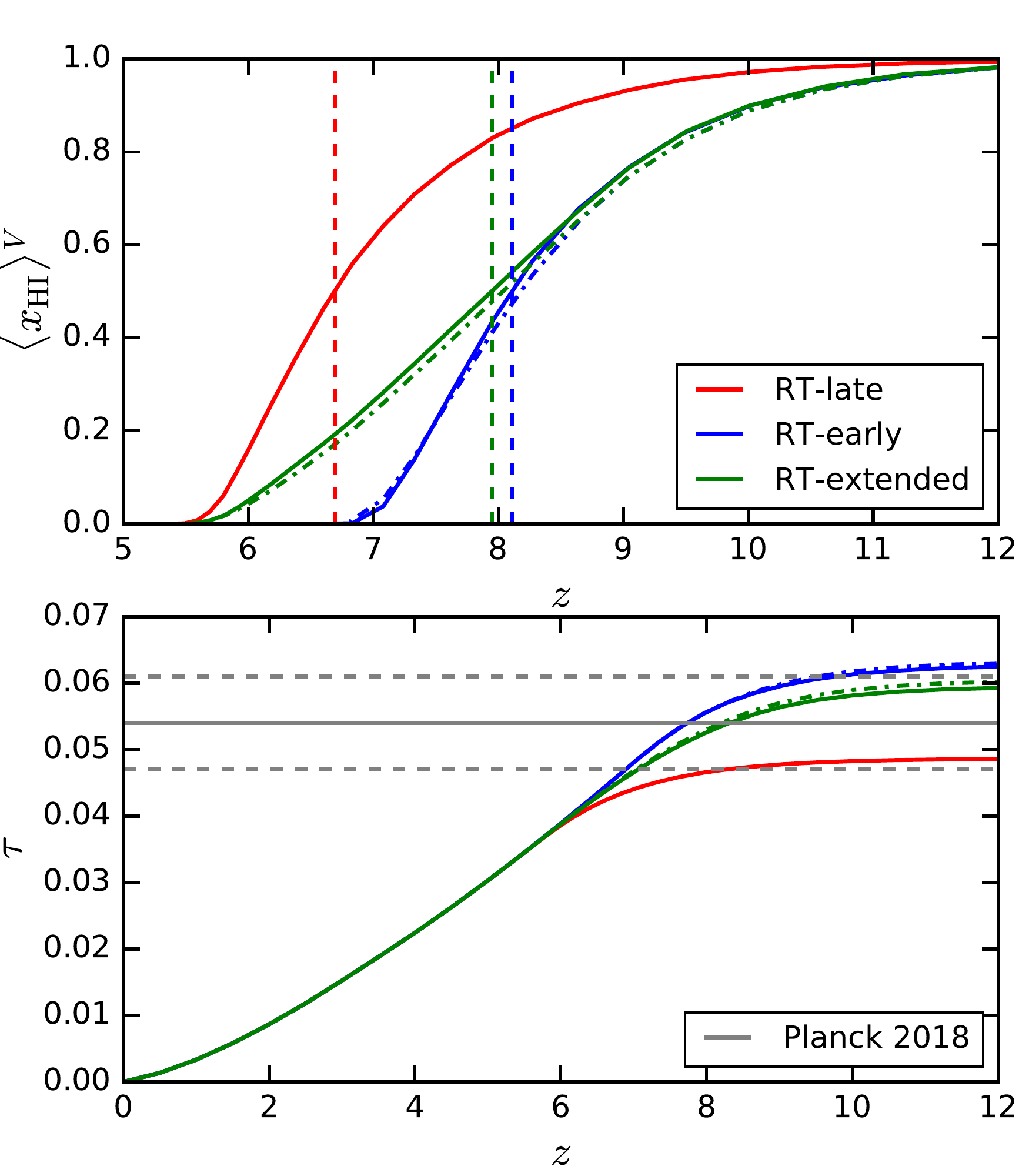}
\caption{Top: the volume-averaged \hi fraction as a function of redshift in all RT simulations. Red, blue, and green colors represent RT-late, RT-early, and RT-extended, respectively. L25n512 and L37.5n768 results are shown by the solid and dot-dashed curves, respectively. The dashed vertical lines illustrate the midpoints of reionization in the L25n512 simulations.
Bottom: the Thomson scattering optical depth in each RT simulation, compared to the \protect\citet{Planck18} observations (gray lines). All of our simulations have Thomson $\tau$ within $\approx1\sigma$ of the observed value.
}
\label{fig:HI-z}
\end{figure}

\begin{table}
\centering
\caption{Summary of simulations, from left to right: simulation name, method, midpoint and endpoint of reionization, the form of $f^{\rm birth}_{\rm esc}$ adopted (if applicable), the stopping redshift.}
\label{tab:sims}
\begin{tabular}{lccccc}
\hline
name & method & $z_{50}$ & $z_{0.001}$ & $f^{\rm birth}_{\rm esc}$ & $z_{\rm stop}$ \\
\hline
L25n512 RT-late & RT & 6.69 & 5.5 & 0.4 & 5.0 \\
L25n512 RT-early & RT & 8.11 & 6.8 & 1.0 & 5.0 \\
L25n512 RT-extended & RT & 7.95 & 5.6 & eq.~\ref{eq:fesc} & 5.0 \\
L25n512 FR-z6.7 & flash & 6.7 & 6.7 & -- & 4.6 \\
L25n512 FR-z8.0 & flash & 8.0 & 8.0 & -- & 4.6 \\
L25n512 FR-z10 & flash & 10 & 10 & -- & 4.6 \\
L25n512 FR-z12 & flash & 12 & 12 & -- & 5.0 \\
L37.5n768 RT-early & RT & 8.18 & 6.8 & 1.0 & 5.0 \\
L37.5n768 RT-extended & RT & 8.04 & 5.7 & eq.~\ref{eq:fesc} & 5.0 \\
\hline
\end{tabular}
\end{table}

Fig.~\ref{fig:HI-z} shows the volume-averaged \hi fraction as a function of redshift (top panel) and the Thomson scattering optical depth to CMB photons (bottom panel) in all RT simulations. Red, blue, and green lines represent RT-late, RT-early, and RT-extended respectively. Solid and dot-dashed lines illustrate L25n512 and L37.5n768 respectively. The reionization midpoints in the L25n512 RT simulations are shown by the dashed lines.
Reionization ends at $z\approx5.5, ~5.6,$ and $6.8$, with mid-points at $z\approx6.7, ~8.0,$ and $8.1$, in RT-late, RT-extended, and RT-early, respectively. These reionization histories result in Thomson scattering optical depths within $\approx1\sigma$ of the observed value of \citet{Planck18}. Because we are interested in the effects of $T$ fluctuations, we do not run RT simulations with earlier reionization partly because $T$ fluctuations hardly remain for the $z\sim5$ power spectrum where the Ly$\alpha$ forest has been measured (and so the effect of earlier reionization can be modeled with less expensive uniform reionization models). The complete list of simulations is given in Table~\ref{tab:sims} with the reionization midpoints and endpoints, when the volume-averaged \hi fraction equals $0.5$ and drops to $\sim10^{-3}$, respectively.

We create mock Ly$\alpha$ forest spectra from the $z=5.4$ and $z=5.0$ snapshots of each simulation (also $z=4.6$ for three FR simulations) using a pixel size of 1~km/s.
The optical depth of each pixel is calculated via integrating $n_{\rm \hi} \sigma_\nu$ along the line-of-sight, where $\sigma_\nu$ is the cross section of Ly$\alpha$ scattering. Each gas cell intersected by a sightline is seen as an ``absorber'' with a certain temperature, peculiar velocity, and Hubble flow across it. The range of integration for each gas cell is determined by the intersection of the sightline with the Voronoi mesh\footnote{See the latest version of \url{https://github.com/sbird/fake_spectra}}.

We generate 5000 sightlines along one axis of the simulation box and calculate the power spectrum of the fractional transmission $\delta_F = F/\langle F \rangle -1$, where $\langle F \rangle$ is the mean transmitted flux.
Following the usual practice, we rescale the optical depth of each pixel to get a desired mean flux in order to compare the simulated power spectrum with observations. We adopt $\langle F \rangle = 0.080$ at $z=5.4$ and $\langle F \rangle = 0.184$ at $z=5.0$ as default (Sec.~\ref{sec:results}), except when fitting the simulated power spectrum to observations (Sec.~\ref{sec:data}). At $z=5.4$, $\langle F \rangle = 0.080$ is nearly twice $\langle F \rangle = 0.046$ used in \citet{Viel13}, but consistent with the recent measurement of \citet{Bosm18}. This value of $\langle F \rangle$ also brings our simulated power spectra into better agreement with the observational data of \citet{Viel13}. At $z=5$, the adopted $\langle F \rangle$ is the same as obtained by \citet{Boera18}. We also show results where the optical depth in different simulations uses the same scaling factor, so all simulations have the same UVB. In Appendix~\ref{sec:mf} we will show that the effects of varying the mean flux on the power spectrum are basically only a shift in its amplitude \citep[see also Fig.~3 of][]{Boera18}, so our conclusions regarding the effects of temperature fluctuations are robust to the choice of $\langle F \rangle$. While using the global mean flux is a standard approach in the literature, the method for calculating the power spectrum should mimic that of reducing real observational data, especially when comparing simulations to observations. We therefore also present an estimate on the differences of using the rolling mean flux to calculate the power spectrum compared to using the global mean in Appendix~\ref{sec:rollingmean}. The rolling mean defines the mean flux locally. For example, \citet{Boera18} uses a $40\, h^{-1}$Mpc boxcar window. We will show that at $z=5.4$, using the rolling mean in an RT simulation raises the small-scale power by $\sim10-20\%$ more than in an FR simulation, while at $z=5.0$ the differences of using the rolling mean are negligible between RT and FR.

Our simulations were chosen to achieve the minimum resolution needed to adequately correct for the effects of resolution, allowing us to achieve the largest box sizes possible. In order to correct for resolution effects, we have run L12.5n256, L12.5n512, L12.5n640, and L12.5n768 simulations with the uniform FG09 UV background. This series allows us to extrapolate to the ``true'' infinite resolution power spectrum using Aitken's delta-squared process \citep{numerical_recipes}. We find that the resolution of L25n512 and L37.5n768 requires $\sim60\%$ and $90\%$ correction to the power spectrum at $k= 0.1$~s/km at $z=5.0$ and $z=5.4$ respectively, and at lower wavenumbers the effect is smaller. L12.5n256 and L12.5n512 RT-extended simulations were also run, showing that the FR and RT simulations roughly need the same amount of resolution correction (and our RT-extended has a much later reionization than the FG09 UV background simulations, showing that the resolution correction is weakly sensitive to temperature). We thus expect to be able to correct resolution effects in our simulations to a fractional accuracy of $\approx 20\%$ at $k= 0.1-0.2$~s/km and even better at lower wavenumbers. In our plots, the observational data points are corrected to mimic the effect of our finite resolution.  Appendix~\ref{sec:reso_corr} discusses the effects of resolution and correction procedure in detail.

\section{Imprints of temperature fluctuations on the power spectrum}
\label{sec:results}

\begin{figure*}
\includegraphics[width=2\columnwidth]{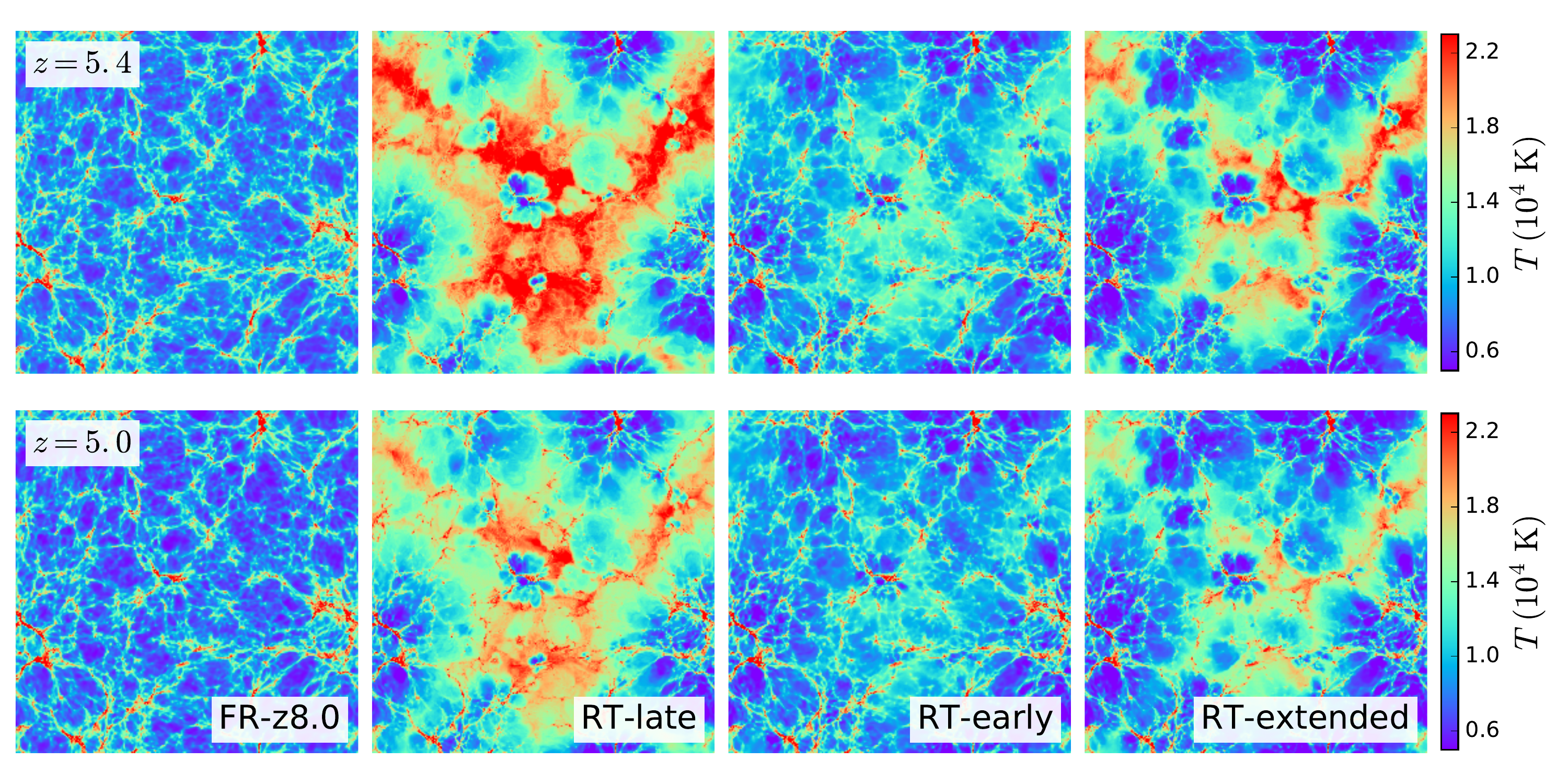}
\caption{A $25\times 25 (h^{-1}{\rm Mpc})^2$ slice showing a $0.5\ h^{-1}{\rm Mpc}$ projection of the post-reionization temperature field at $z=5.4$ (top panels) and $z=5$ (bottom panels) in the L25n512 simulations. From left to right, the slices are taken from FR-z8.0, RT-late, RT-early, and RT-extended, respectively. The temperature fields in FR-z8.0 largely trace the underlying density field, while those in the RT simulations can show significant large-scale scatter owing to patchy reionization.}
\label{fig:vis_T}
\end{figure*}

\begin{figure*}
\includegraphics[width=2\columnwidth]{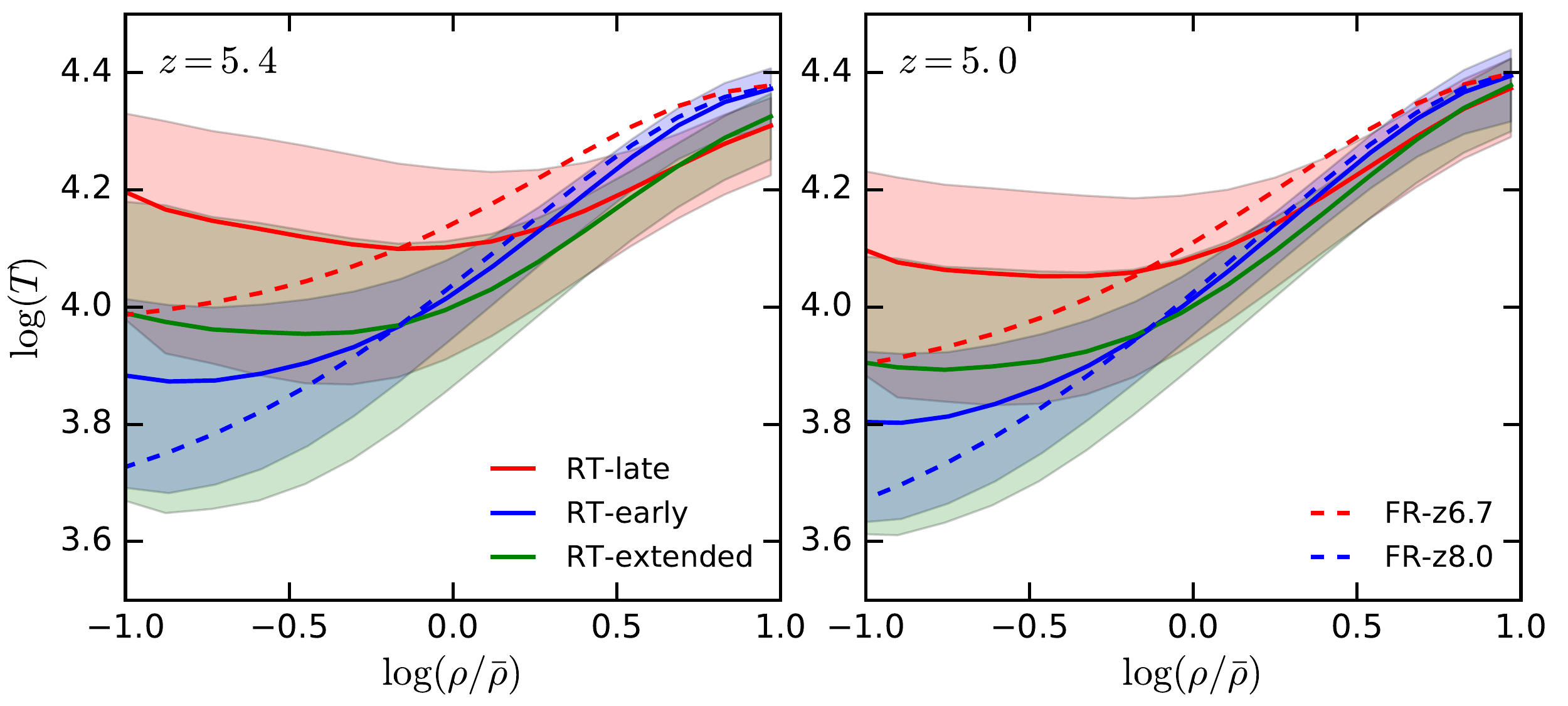}
\caption{Mean temperature--density relations in the L25n512 RT (solid lines) and FR (dashed lines) simulations at $z=5.4$ (left panel) and $z=5.0$ (right panel). Red, blue, and green colors represent RT-late and FR-z6.7, RT-early and FR-z8.0, and RT-extended, respectively. Shaded regions encapsulate the $1\sigma$ scatter in the temperature, which are only shown for the RT simulations as this scatter is small in the FR simulations. RT-late produces the highest IGM temperature. RT-extended has the largest $T$ fluctuations at $z\sim5$. We note that at $z\sim5$, the gas density range that contributes most to the transmission is $\rho/\bar{\rho}\sim0.2-0.4$ \protect\citep{FO09}.}
\label{fig:temp-dens}
\end{figure*}

We first present an overview of temperature fluctuations in different simulations and the resulting Ly$\alpha$ forest flux power spectra.
Our simulations differ in the amplitude of temperature fluctuations. Fig.~\ref{fig:vis_T} shows projections of the temperature field at $z=5.4$ and $z=5.0$ in the FR-z8.0 and L25n512 RT simulations in a $25\times25\times0.5\ (h^{-1}{\rm Mpc})^3$ volume. The temperature field in FR-z8.0 traces the structures of the underlying density field, whereas the gas temperatures in the RT simulations show different amounts of scatter due to patchy reionization.
Fig.~\ref{fig:temp-dens} considers the relationship between $T-\rho$ in the L25n512 RT and FR simulations. Solid and dashed curves show the mean $T-\rho$ relations in the RT and FR simulations, respectively. Red, blue, and green represent RT-late and FR-z6.7, RT-early and FR-z8.0, and RT-extended, respectively. The shaded regions in Fig.~\ref{fig:vis_T} encapsulate the $1\sigma$ scatter in $T$ of the RT simulations. The scatter in $T$ at $\rho/\bar{\rho}\sim0.1-0.5$ is $2-3$ times smaller in RT-early than in RT-late, because gas in the former was ionized the earliest and hence has the longest time to cool. Conversely, RT-late produces the highest gas temperatures. The largest $T$ fluctuations are seen in RT-extended since it combines gas that was ionized early with gas ionized late. We find that the larger box L37.5n768 RT simulations have almost the same $T-\rho$ relations as the L25n512 RT simulations. The gas density range that contributes most to the transmission (and hence is most important for the forest power spectrum) is $\rho/\bar{\rho}\sim0.2-0.4$ at $z\sim5$ \citep{FO09}, where each RT simulation produces $<0.1$ dex higher temperatures than its FR counterpart. The mean $T-\rho$ relations in voids are also steeper in FR than in RT.
The temperature range that all of our simulations cover is $T_0\approx9,000-13,000$~K at the mean density at $z\sim5$.\footnote{We note that the lowest value of $T_0$ at $z\sim5$ that our simulations reach is $\approx9000$~K in FR-z10 and FR-z12, about $\sim1000$~K higher than simulations in the literature using the \citet{HM12} UV background. This is caused by the harder spectrum of the FG09 UVB model and the photoionization of HeII adding $\approx1000$~K to the IGM temperature by $z=5$ in this model. The HeII heating increases to lower redshifts, leading to a rather flat $T_0$ evolution.}

\begin{figure*}
\includegraphics[width=2\columnwidth]{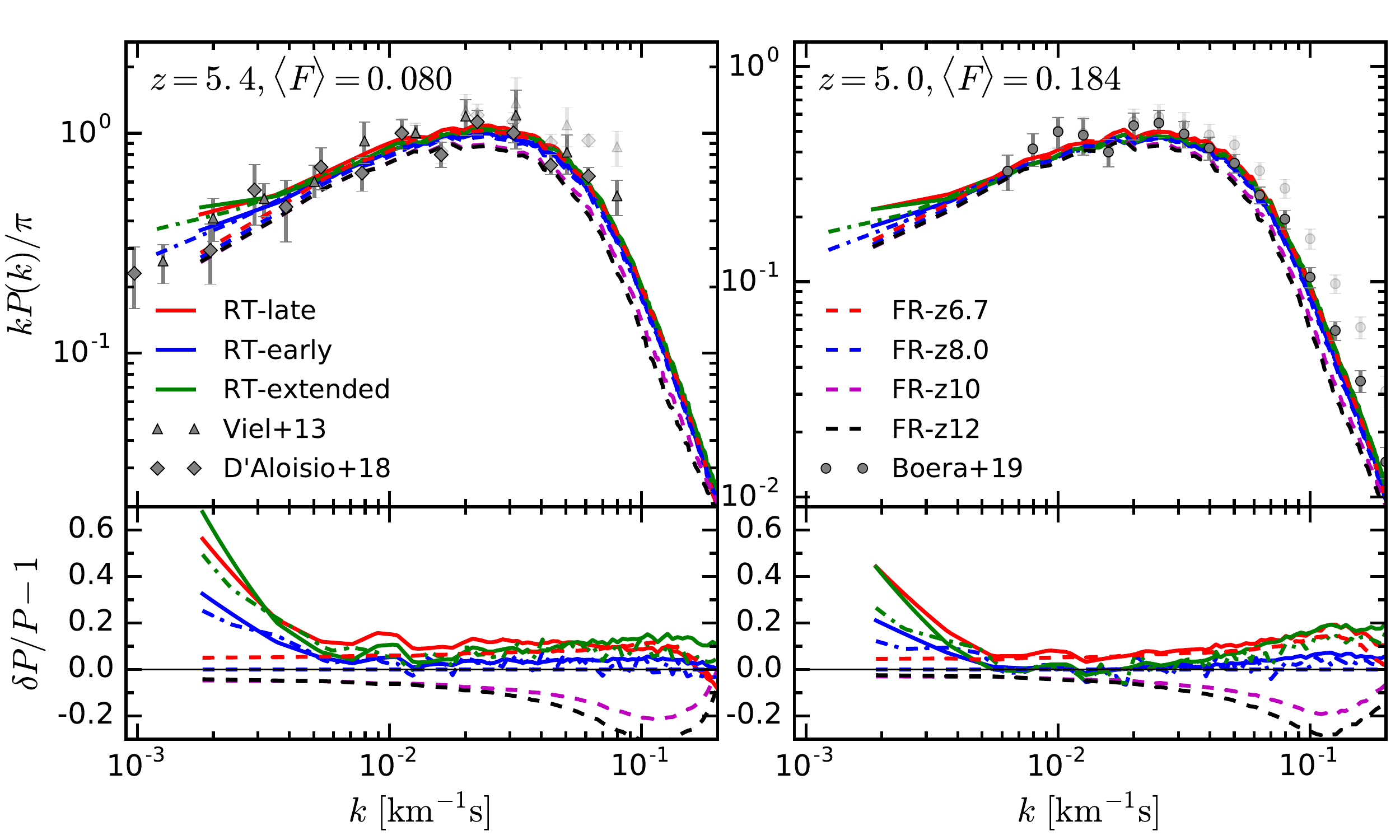}
\caption{Power spectra in the L25n512 RT (solid lines), FR (dashed lines) and  L37.5n768 RT (dot-dashed lines) simulations at $z=5.4$ (left panel) and $z=5$ (right panel). All power spectra are normalized to the same mean flux, $\langle F \rangle$, given in the top of each top panel. Observational measurements from \protect{\citet[][gray triangles]{Viel13}}, \protect{\citet[][gray diamonds]{DAlo18fluc}}, and \protect{\citet[][gray circles]{Boera18}} have been corrected to our finite simulations' resolution, with the original uncorrected measurements shown in light gray.
Bottom panels show the fractional differences of the power spectra in each simulation with respect to those in FR-z8.0. The differences among power spectra with reionization midpoints $z\le8$ are within $\sim20\%$ at small and intermediate scales ($0.005 \lesssim k \lesssim 0.2$~s/km), while at large scales the L25n512 RT simulations generate up to $\sim20-60\%$ extra power than FR-z8.0 at $k=0.002$~s/km.
}
\label{fig:powerspectrum}
\end{figure*}

Fig.~\ref{fig:powerspectrum} presents the power spectra in the L25n512 RT (solid lines), FR (dashed lines), and L37.5n768 RT (dot-dashed lines) simulations at $z=5.4$ (left panels) and $z=5.0$ (right panels), compared to the resolution-corrected observational data of \citet[][gray triangles]{Viel13} and \citet[][gray diamonds]{DAlo18fluc} at $z=5.4$, and that of \citet[][gray circles]{Boera18} at $z=5.0$. The uncorrected observations are shown as light gray symbols. All power spectra are normalized to the same mean flux: $\langle F \rangle = 0.08$ at $z=5.4$ and $\langle F \rangle = 0.184$ at $z=5.0$. % Ten percent changes in the mean flux lead to  $6-10\%$ shifts in the power spectrum, with the effect mostly in the overall amplitude. This represents errors introduced by the unknown mean flux \citep[see also][]{Onor17ps}.
The bottom panels of Fig.~\ref{fig:powerspectrum} show the fractional differences of the power spectra with respect to those in FR-z8.0.
Two features are most prominent in this comparison: large-scale excess power in the RT simulations at $k\lesssim0.004$~s/km that increases to $20-60\%$ at $k=0.002$~s/km, and the $\lesssim20\%$ ($\lesssim40\%$) differences in the small and intermediate-scale power in simulations with reionization midpoints $z\le8$ ($z\le12$).
In the following, we will use two tests to investigate the physical origin of the large-scale (Sec.~\ref{sec:largescale}) and small-scale (Sec.~\ref{sec:thermalbroadening} and \ref{sec:pressuresmoothing}) features, focusing in particular on the role of $T$ fluctuations.

\begin{figure*}
\includegraphics[width=2\columnwidth]{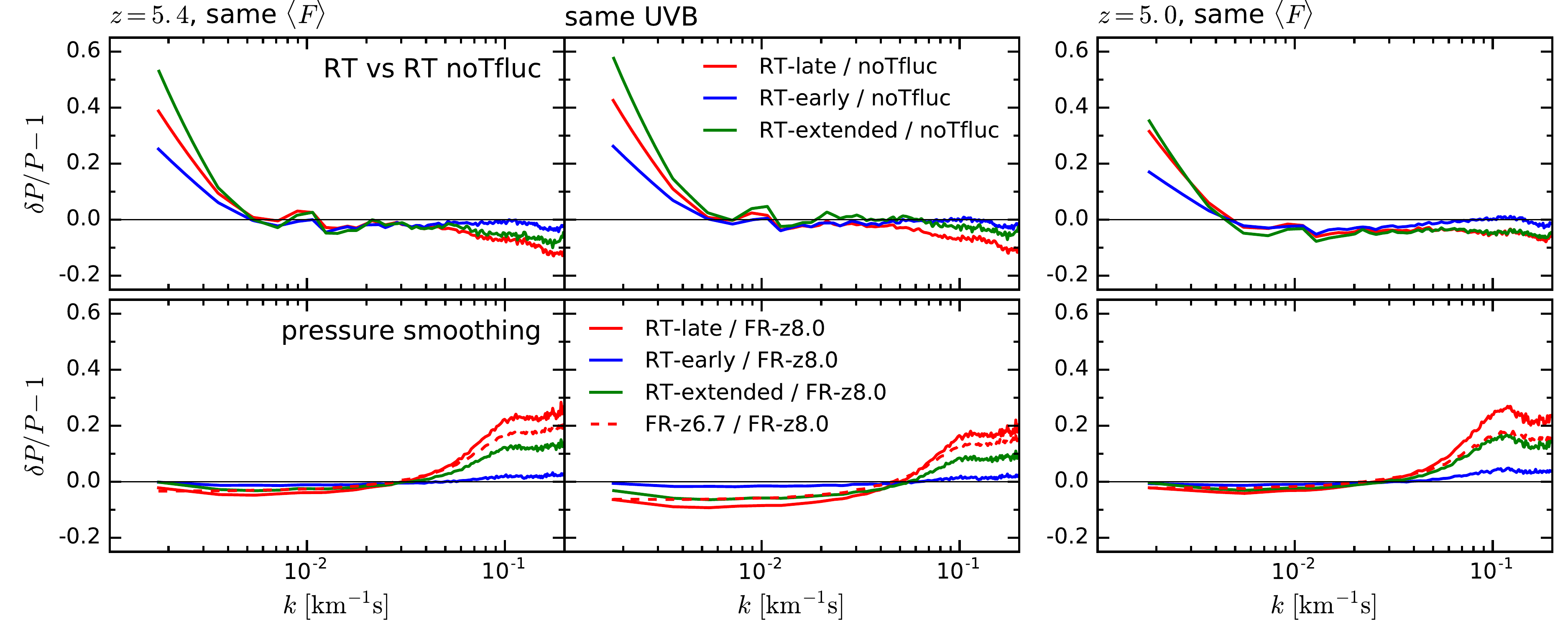}
\caption{Fractional differences of the L25n512 power spectra in various tests.
Results at $z=5.4$ are shown in the left and middle columns, with the power spectra normalized to the same mean flux and to the same UVB, respectively. The right column presents results at $z=5.0$ where the power spectra are normalized to the same mean flux.
Top panels: power spectra in the RT simulations compared to those where $T$ fluctuations are artificially removed by placing gas cells onto their mean temperature--density relations. Temperature fluctuations are found to suppress the small-scale power by $\lesssim10\%$, in addition to bringing large-scale extra power.
Bottom panels: each power spectrum in a simulation is compared to the one in FR-z8.0, with the gas cells in all simulations put onto the temperature--density relations in FR-z8.0 to separate the pressure smoothing effects.
}
\label{fig:residual}
\end{figure*}

\subsection{Imprints of large-scale $T$ fluctuations on the low-$k$ power}
\label{sec:largescale}

We first examine how $T$ fluctuations leave imprints on the low-wavenumber end of the power spectrum, where the L25n512 RT simulations show $\sim20-60\%$ excess power compared to FR-z6.7 and FR-z8.0 at $k=0.002$~s/km.
While the FR simulations provide a toy model of no $T$ fluctuations, the effects of $T$ fluctuations can be investigated further by artificially removing them.
We do so by putting gas cells with $\log(\rho/\bar{\rho})\in[-1,1]$ and $T<10^5$~K in each L25n512 RT simulation onto their mean $T-\rho$ relations at $z=5.4$ and $z=5.0$ and calculating the corresponding power spectra. This rescaling not only affects the amount of thermal broadening but also the \hi fraction. Assuming photoionization equilibrium, which should be an excellent approximation, the \hi fraction of a gas cell is proportional to the recombination rate $\alpha(T)$ at temperature $T$. Thus when changing the gas temperature to $T'$, we scale the gas \hi fraction by $\alpha(T')/\alpha(T)$ to remove inhomogeneities in the \hi fraction. Since this operation is performed in post processing such that the gas distribution is unchanged, the pressure smoothing is the same between the two power spectra and so this isolates other thermal effects.

The top panels of Fig.~\ref{fig:residual} show the fractional differences of the original power spectra compared to those rescaled to the mean $T-\rho$ relations. 
The left and middle columns show results at z = 5.4, where the power spectra are normalized to the same mean flux and the same UVB, respectively. Results at z = 5.0 are presented in the right column, with the power spectra normalized to the same mean flux. Rescaling to the same mean flux is how models are compared to observations (since the UVB amplitude is unknown), while normalizing to the same UVB compares on a more physical footing. 
Similar to the comparison with FR-z6.7 and FR-z8.0, extra power is seen at $k\lesssim0.004$~s/km. At $z=5.4$, RT-extended, RT-late, and RT-early produce $\sim50\%$, $40\%$, and $20\%$ more power at $k=0.002$~s/km than their no $T$ fluctuation counterpart, respectively.
Large-scale coherent $T$ fluctuations from thermal broadening and \hi fraction variations thus bring in extra large-scale power. The larger the $T$ fluctuations are (RT-extended and RT-late), the more power there is at low $k$. 
This large-scale excess power decreases with time as $T$ fluctuations fade away, dropping by $\sim10\%$ from $z=5.4$ to $z=5.0$.

We note that at low $k$, RT-late and RT-extended appear to overshoot the power found in the $z=5.4$ measurements, especially if the trend in each power spectrum is extrapolated to lower $k$ (Fig.~\ref{fig:powerspectrum}). However, the L37.5n768 RT-extended and RT-early simulations have $\sim10-20\%$ less power at $k=0.002$~s/km than L25n512 RT-extended and RT-early. These differences could owe to sample variance or the artificial suppression of structures of our small box simulation.

\subsection{Imprints of $T$ fluctuations on the high-$k$ power excluding pressure smoothing}
\label{sec:thermalbroadening}

Turning to higher wavenumbers, the top panels (in all three columns) of Fig.~\ref{fig:residual} show that $T$ fluctuations suppress power at $k\gtrsim0.1$~s/km by $\sim5-10\%$ compared to the no $T$ fluctuation case, especially in RT-late and RT-extended\footnote{We note, however, that the small-scale effects of $T$ fluctuations also depend on the method of removing $T$ fluctuations. Using the median $T-\rho$ relation instead of the mean enlarges the suppression by $\sim5-10\%$, because the median $T$ is generally lower than the mean.}.
This suppression is in disagreement with previous works that argued the high-$k$ end of the power spectrum is dominated by contributions from cold regions resulting in more small-scale power \citep[e.g. equation 2 in][]{McQu11, Hui17}.
Our findings can be explained with a simple toy model. Assume that the universe is composed of $N$ uncorrelated patches of size $L$, and that each patch $i$ has a local mean flux $\langle F \rangle_{\rm local,i}$.  The average of $\langle F \rangle_{\rm local,i}$ yields the global mean flux, $\langle F \rangle_{\rm global}$. Denoting $\mathcal{F}$ as the Fourier transform operator, the global (or observed) power spectrum at $k\gg L^{-1}$ is approximately the average of all the local power spectra:
\begin{align*}
 L P_{\rm global}(k) &\approx \frac{1}{N}\sum_i \Big| \mathcal{F} \left( F_i/ \langle F \rangle_{\rm global}-1 \right) \Big|^2, \\
&\approx \frac{1}{N}\sum_i \underbrace{\left( \langle F \rangle_{\rm local,i}/\langle F \rangle_{\rm global} \right)^2}_{w_i} \Big| \mathcal{F} \left( F_i / \langle F \rangle_{\rm local,i}-1 \right) \Big|^2, \\
&= \frac{1}{N}\sum_i w_i  ~L P_{\rm local,i}(k),
\end{align*}
where the approximation from the first line to the second holds for $k\neq0$.  The global power spectrum is therefore a weighted average of the local power spectrum, with the weights roughly given by $w_i \equiv (\langle F \rangle_{\rm local,i}/\langle F \rangle_{\rm global})^2$.
Because the hotter regions have higher local mean fluxes than the colder ones (the \hi fraction scales as $\sim T^{-0.7}$), the local power spectra of the hotter patches contribute more heavily to the global power spectrum. Therefore, the high-$k$ end of the global power spectrum at relevant wavelengths receives a larger contribution from higher temperatures, resulting in a stronger `exponential' suppression in the small-scale power.

\begin{figure}
\centering
\includegraphics[width=\columnwidth]{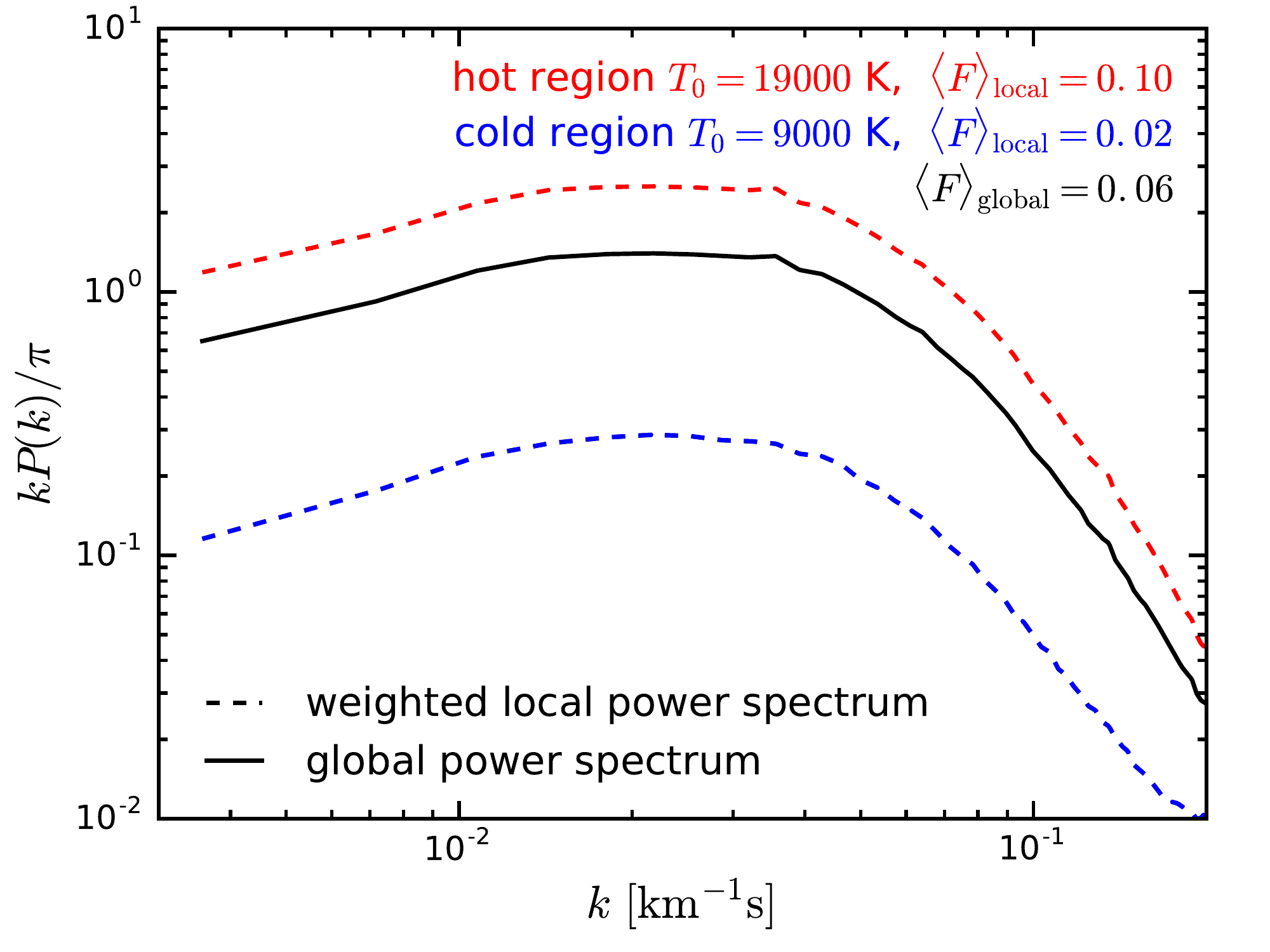}
\caption{An illustration of why the global power spectrum is weighted more heavily towards hotter regions. Blue and red dashed lines show the mean flux-weighted local power spectra at $z=5.4$ in two FR simulations reionized at $z=10.5$ and $5.5$ respectively. The weights are given by $\left( \langle F \rangle_{\rm local}/\langle F \rangle_{\rm global} \right)^2$. The values of $\langle F \rangle_{\rm local}$, $\langle F \rangle_{\rm global}$, and gas temperatures at mean density ($T_0$) are given in the top right corner of the figure. The solid line is the global power spectrum, which is the average of the mean flux-weighted local power spectra assuming the universe is $50\%$ hot and $50\%$ cold. Hotter regions have higher $\langle F \rangle_{\rm local}$, hence dominating the contribution to the global power spectrum.}
\label{fig:local_powerspectrum}
\end{figure}

We illustrate this idea in Fig.~\ref{fig:local_powerspectrum}, showing the mean flux-weighted local power spectra, $w_i P_{\rm local,i}(k)$, of two extreme FR simulations at $z=5.4$.
One is reionized at $z=5.5$ (red dashed) and the other at $z=10.5$ (blue dashed) to mimic hot and cold regions. The former has $\langle F \rangle_{\rm local} = 0.10$ and $T_0 = 19,000$~K at $z=5.4$, where $T_0$ is the gas temperature at mean density, and the latter has $\langle F \rangle_{\rm local} = 0.02$ and $T_0 = 9000$~K. Assuming that the universe is $50\%$ hot and $50\%$ cold, $\langle F \rangle_{\rm global}=0.06$. With a much smaller local mean flux, the power spectrum of the cold region becomes much lower in amplitude than that of the hot one due to the lower weight of the former. The small scale features of the global power spectrum (black solid line) are thus mostly determined by the hot region. This explains the $\lesssim10\%$ suppression of small-scale power by $T$ fluctuations seen in the top panels of Fig.~\ref{fig:residual}.

Our discussion in this section may not apply to all observations as some of them, such as \citet{Boera18}, adopt a rolling mean to calculate the overdensity in the normalized flux. A rolling mean makes sightlines with lower mean fluxes more important because it defines the mean flux locally. This could undermine our argument that hotter regions dominate the power \emph{if} the coherence length of temperature fluctuations is comparable or larger than the length used to define the rolling mean. In Appendix~\ref{sec:rollingmean}, we find that the suppression of the small-scale power due to $T$ fluctuations is almost eliminated at $z=5.4$ if the power spectrum is calculated using the rolling mean (i.e. the left and middle panels of the top row of Fig.~\ref{fig:residual}). At $z=5.0$, using the rolling mean gives very similar results as the global mean, but the suppression of the high-$k$ power by $T$ fluctuations is only at $\sim5\%$ level, therefore negligible. 
A larger simulations box, allowing larger coherence lengths, may also result in a bigger effect. However, there is an indication that we are overestimating the effect of a rolling mean. We estimate that the rolling mean raises the amplitude of the power spectrum at $z=5.0$ by $\sim10\%$, but \citet{Boera18} showed that using a $40\, h^{-1}$Mpc boxcar window recovers the power on all scales as using the global mean (see Appendix~\ref{sec:rollingmean} for details). More detailed work is therefore needed to understand the differences between our findings and those of \citet{Boera18}, and how $T$ fluctuations affect the small-scale power.%\matt{one may worry that, since none of these simualtions are capturing coherence of large-scale opacity fluctuations, this may bias determinations from high k power spectrum} 

\begin{figure}
\centering
\includegraphics[width=\columnwidth]{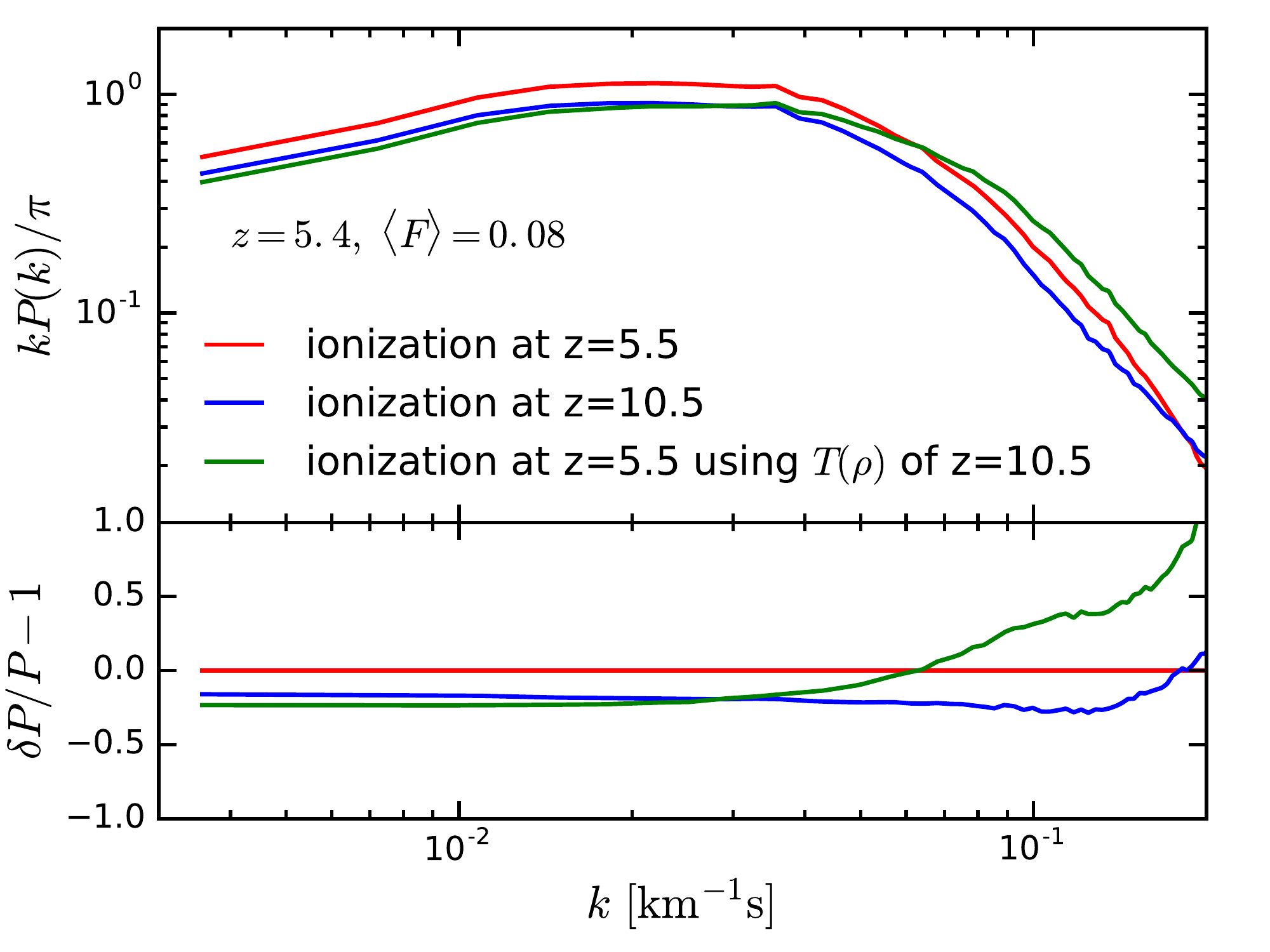}
\caption{$z=5.4$ power spectra in FR simulations ionized at $z=5.5$ (red) and $z=10.5$ (blue), normalized to the same mean flux of $0.08$. Green line represents power spectrum of the $z=5.5$ FR simulation, with the gas placed onto the $T-\rho$ relation of the $z=10.5$ FR simulation. Bottom panel shows fractional differences of the power spectra with respect to the one from the $z=5.5$ FR simulation. Pressure smoothing in the $z=10.5$ FR simulation suppresses power by $\sim50\%$ at $k\sim0.01$~s/km, significantly cancelling the suppression of small-scale power by thermal broadening in the $z=5.5$ FR simulation.}
\label{fig:diff}
\end{figure}

Interestingly, the shapes of the two local power spectra in Fig.~\ref{fig:local_powerspectrum} are very similar except at the highest wavenumbers, despite the $10,000$~K difference in $T_0$. Fig.~\ref{fig:diff} further illustrates this, with the power spectra in the two FR simulations normalized to the same mean flux of $0.08$ at $z=5.4$ (red and blue lines). These power spectra differ by $\lesssim30\%$ on all scales. We note that there is almost no pressure smoothing at $z=5.4$ in the FR simulation ionized at $z=5.5$, so the power spectrum of this simulation is only shaped by thermal broadening. On the other hand, gas in the simulation ionized at $z=10.5$ has a long time to relax. The green line in Fig.~\ref{fig:diff} represents power spectrum in the $z=5.5$ FR simulation, if the gas follows the $T-\rho$ relation of the $z=10.5$ FR simulation. Comparison to the power spectrum of the $z=10.5$ FR simulation shows a $\sim50\%$ suppression of power at $k\sim0.1$~s/km due to pressure smoothing effects in the latter simulation. This indicates a strong cancellation between thermal broadening and pressure smoothing effects, resulting in very similar shapes of the power spectra in simulations ionized at very different times. We explore the effects of pressure smoothing in detail below.

\subsection{Effects of pressure smoothing on the high-$k$ power}
\label{sec:pressuresmoothing}

We now discuss the pressure smoothing effects in different simulations, which affect the high-wavenumber end of the power spectrum in addition to thermal broadening.
Specifically, sound waves have had more time to smooth out the gas density field in regions that are ionized earlier than later. Earlier reionization thus results in more pressure smoothing of the gas. We will focus especially on the cancellation between the effects of thermal broadening and pressure smoothing.
To examine pressure smoothing effects, we place the gas cells in each L25n512 simulation (except FR-z10 and FR-z12) with $\log(\rho/\bar{\rho})\in[-1,1]$ and $T<10^5$~K onto the $T-\rho$ relations in FR-z8.0 at $z=5.4$ and $z=5.0$ and calculate the corresponding power spectra. The \hi fraction of the gas cells is again scaled by the ratio of the recombination rates at the new and old temperatures: $\alpha(T')/\alpha(T)$. Since all simulations are forced to have the same $T-\rho$ relation, differences in the power spectra only reflect differences in pressure smoothing effects.
The bottom panels of Fig.~\ref{fig:residual} shows the fractional differences of the resulting power spectra in the L25n512 RT and FR-z6.7 simulations compared to those in FR-z8.0. At both redshifts, more pressure smoothing suppresses the small-scale power at $k\gtrsim0.04$~s/km, consistent with the results of previous studies \citep[e.g.][]{Onor17ps, Boera18}.

We first examine the pressure smoothing in the RT simulations and illustrate its degeneracy with thermal broadening on the small-scale power. As found in the bottom panel of Fig.~\ref{fig:powerspectrum}, at small and intermediate scales ($0.005 \lesssim k \lesssim 0.2$~s/km), the differences among the power spectra in the RT simulations are within $\sim20\%$, although RT-late produces a factor of $\sim1.5$ higher mean temperatures in gas with $\rho/\bar{\rho}\sim0.2-0.4$ than RT-early and RT-extended. This appears inconsistent with previous works where simulations with higher temperatures harbor more of a small-scale suppression owing to  thermal broadening \citep[e.g.][]{Peep10, Nasir16, Onor17ps, Boera18}. However, in the bottom panels of Fig.~\ref{fig:residual}, less pressure smoothing in RT-late and RT-extended generates $\sim20\%$ and $\sim10\%$ more power at $k\sim0.1$~s/km than RT-early, respectively. The combination of more pressure smoothing and reduced thermal broadening (from lower temperatures) in RT-early therefore produces similar small-scale power as the pair of less pressure smoothing and higher temperatures in RT-late. Table~\ref{tab:lp} demonstrates this further, where we list the IGM temperatures at the mean density ($T_0$) and the pressure smoothing scales ($\lambda_p$) in each simulation at $z=5.0$. The $\lambda_p$ values are calculated following \citet{Kulk15}.\footnote{We note that the IGM is very cold ($\sim10-100$~K) before reionization in our simulations, but X-ray heating likely heats the IGM up to $\sim1000$~K long before reionization \citep[e.g.][]{Oh03} and contributes to the amount of pressure smoothing.  Suppose X-ray heating starts at $z=20$ heating the Universe to $1000$~K, and reionization occurs instantaneously at $z_{\rm re}$ with the temperature jumping to $10,000$~K. The contributions to $1/k^2_F$ from the two epochs can therefore be estimated using equation 11 of \citet{GH98}, where $k_F$ is the filtering scale. We find that in this toy model, the contribution from X-ray heating becomes comparable or larger at $z=5.0$ than the contribution from reionization if $z_{\rm re}\le6$. If $z_{\rm re}>7$, the contribution from X-ray heating is at least a factor of 5 smaller than the contribution from reionization. However, this estimate depend on the assumed temperatures as well. If the temperature boost during X-ray heating is much lower than $1000$~K \citep[e.g. of order 10~K as suggested by][]{Eide18}, the X-ray heating era would contribute much less. Moreover, for our extended reionization histories, the redshift at which half the gas is ionized is $z\geq 6.7$, even though reionization ends very late at $z\approx 5.5$ in two of the simulations. Therefore the effects of the X-ray heating era on the total pressure smoothing can likely be ignored.} There is thus a substantial cancellation between pressure smoothing and thermal broadening effects, which reduces the high-$k$ differences in power among the RT simulations (and also between FR-z6.7 and FR-z8.0, and the two FR simulations shown in Fig.~\ref{fig:diff}). This effect is missed in post-processing hydrodynamic simulations with RT, where pressure smoothing is not self-consistently modeled.

This massive cancellation between thermal broadening and pressure smoothing effects is also noticed in previous works \citep[e.g. Fig.5 of][]{Nasir16, Boera18}, but is more prominent for the more physically motivated thermal histories in our simulations than those in the literature.
Previous works often use a uniform UV background that is turned on at some reionization redshifts with the photoheating rates scaled by different factors to vary the IGM temperature \citep[e.g,][]{Peep10, Nasir16, Onor17ps, Boera18}. This achieves a very broad parameter space of temperature and pressure smoothing, where some combinations of parameters are unphysical. For our more physically motivated models which only cover a narrow subset of this large parameter space, the degeneracy between thermal broadening and pressure smoothing is more evident. %Thus when comparing simulations with different IGM temperatures, those with higher temperatures often have more pressure smoothing as well \citep[e.g. Fig.~3 in][]{Boera18}, leading to larger cut-offs in the small-scale power.
%Therefore for our more physically motivated  ionization models, the shape of the power spectrum is only mildly affected by temperature.
We also note that the effects of pressure smoothing is more prominent at the redshifts we consider. At lower redshifts $z\sim3$, \citet{Peep10} showed that thermal broadening dominates over pressure smoothing in shaping the small-scale power. Since the thermal broadening scale and the Jeans scale have the same scaling with redshift \citep[Fig.B.4 of][]{Irsic17a}, but the latter also scales as $\rho^{-1/2}$, the pressure smoothing scale could be comparable to the thermal broadening scale at $z\sim5$, when most of the transmission is provided by gas with $\rho/\bar{\rho}\sim0.2-0.4$. At lower redshifts, the thermal broadening scale dominates so the effects of pressure smoothing are not as important.

We finally compare the pressure smoothing between RT and FR simulations, which helps explain the $\lesssim20\%$ differences in the small and intermediate-scale power in RT and FR simulations with the same reionization midpoint (see the bottom panels of Fig.~\ref{fig:powerspectrum}; also found by \citealt{Onor18}).
In the bottom panels of Fig.~\ref{fig:residual}, the similarity of pressure smoothing in RT-late and FR-z6.7 leads to better than $5\%$ agreement between their power spectra at $k\gtrsim0.1$~s/km. Similar differences are present for the RT-early and RT-extended, for which the midpoint is approximated by the FR-z8.0 reference simulation. Therefore when the midpoint of reionization is fixed, an RT simulation has comparable pressure smoothing as an FR simulation. This is also illustrated in Table~\ref{tab:lp}, where the $\lambda_p$ values are similar between an RT simulation and its FR counterpart with the same reionization midpoint.
In addition, thermal broadening effects are also similar between each RT and FR pair, since gas with $\rho/\bar{\rho}\sim0.2-0.4$ in each RT simulation is only $\sim0.1$ dex higher in temperature than in the corresponding FR simulation. Moreover, as shown in \S~\ref{sec:thermalbroadening}, $T$ fluctuations only introduce  $\lesssim10\%$ more suppression in the power at $k\gtrsim0.1$~s/km.
These factors thus combine to give similar amounts of small-scale power in RT and FR simulations with the same reionization midpoint.

\begin{table}
\centering
\caption{Temperatures at mean density ($T_0$, second column) and the pressure smoothing scales ($\lambda_p$, third column, in units of comoving kpc$/h$) in all simulations at $z=5.0$. The corresponding values at $z=5.4$ are very similar.}
\label{tab:lp}
\begin{tabular}{lcc}
\hline
name & $T_0^{z=5.0}$[K] & $\lambda_p^{z=5.0}$[ckpc$/h$] \\
\hline
L25n512 RT-late & 12,100 & 51 \\
L25n512 RT-early & 10,400 & 59 \\
L25n512 RT-extended & 10,100 & 54 \\
L25n512 FR-z6.7 & 12,900 & 56 \\
L25n512 FR-z8.0 & 10,500 & 62 \\
L25n512 FR-z10 & 9,600 & 67 \\
L25n512 FR-z12 & 9,400 & 70 \\
L37.5n768 RT-early & 10,400 & 59 \\
L37.5n768 RT-extended & 10,100 & 55 \\
\hline
\end{tabular}
\end{table}

\section{Discussions}
\label{sec:discussions}

\subsection{Comparison to other works}
\label{sec:comparison}

Previous RT simulations of reionization have also been used to study the high-redshift Ly$\alpha$ forest, in particular \citet{Onor18} and \citet{Keat18}. \citet{Onor18} performed a series of cosmological hydrodynamic simulations of $40\, h^{-1}$ Mpc box size, coupled with semi-analytic models of reionization. Our findings agree with theirs that the RT and FR simulations produce similar small and intermediate-scale power in the Ly$\alpha$ forest when the midpoint of reionization is fixed. At $z\sim5$, the L37.5n768 RT-early simulation generates a comparable excess of large-scale power as their IR-C simulation, which shows a similar reionization history (and uses a similar minimum halo mass for producing ionizing photons). The L37.5n768 RT-extended simulation gives $\sim10-20\%$ more power near $k=0.001$~s/km than their most extended reionization (IR-B) simulation with the highest large-scale power, likely due to a more extended and late-ending reionization history.
The wide distributions of post-I-front temperatures in our simulations, compared to the uniform temperature boost at reionization as assumed in \citet{Onor18}, could also enlarge temperature fluctuations at $z\sim5$. Although our simulation outputs do not contain enough information for an estimate of the post-I-front temperatures (e.g. the redshifts at which the gas cells are ionized), \citet{DAlo18Ifront} suggests that they range from being mostly $\lesssim20,000$~K at the beginning of reionization to $\sim25,000-30,000$~K near the end of reionization. A single $20,000$~K temperature boost at reionization would therefore miss a high-temperature tail given a post-reionization redshift \citep[see Fig.10 and 11 of][]{DAlo18Ifront}. This effect is more relevant for later-ending reionization models where temperature fluctuations have not had enough time to fade away.
The simulations in \citet{Keat18} that were run in post-processing with RT are not able to capture pressure smoothing effects, which we find to cancel much of the effect from thermal broadening. Our simulations also span a broader parameter space of the endpoint and duration of reionization than theirs. We find that their power spectra differs only by $\lesssim15\%$ from ours at $k\in[0.004,0.1]$~s/km when normalized to the same mean flux, which we suspect is a coincidence due to different reionization histories and different levels of numerical convergence of the power spectrum.

Previous studies also shed light on whether our results should be converged. Many studies have found that $>100$ Mpc simulation boxes are required to correctly capture the large clustering scales of the ionizing sources and the resulting ionized regions subtending tens of comoving Mpc or more \citep[e.g.][]{Bark04, Iliev06}. The 21cm signal from reionization is converged at the tens of percent level only in such large boxes. Because the Ly$\alpha$ forest samples skewers rather than the 3D volume, similar convergence may be achieved in smaller boxes. However, the modest differences in the Ly$\alpha$ forest power spectrum between between the L25n512 and L37.5n768 simulations could arise just from standard sample  variance (suggesting that the L37.5 is essentially converged on overlapping scales)  or could indicate a more heinous enhancement at the box scale (indicating that a larger box would result in a further reduction). Thus, we caution box size effects as a major potential caveat in our estimates for the enhancement in the large-scale power from reionization.

\subsection{Implications for interpreting current observations}
\label{sec:data}

\begin{table*}
\caption{$\chi^2$ values calculated for each simulated power spectrum with the \citet{Viel13} data at $z=5.4$, and the \citet{Boera18} data at $z=5.0$ and $z=4.6$. Following \citet{Viel13}, we only use the 7 data points with $-2.3\le\log_{10}k({\rm s/km})\le-1.1$. For the latter two calculations, we list two sets of $\chi^2$ values, one evaluated using 13 data points with $k\le0.1$~s/km (second column), and the other using all 16 data points with $k\le0.2$~s/km (fourth column; other columns in these tables are computed for $k\le0.1$~s/km). We also show the scaling factor of the FG09 UVB at minimum $\chi^2$ (third column). For the $z=5.0$ table, the last column gives $P=\exp\left(-\Delta\chi^2/2\right)$, where $\Delta\chi^2$ is the $\chi^2$ difference of each simulation with respect to FR-z6.7. In the $z=4.6$ table, the fifth column lists the total $\chi^2_{\rm tot}$ by summing up the $\chi^2$ values of all redshift bins. The last column shows $P=\exp\left(-\Delta\chi^2_{\rm tot}/2\right)$, with the $\chi^2_{\rm tot}$ differences taken with respect to FR-z6.7. For the two L37.5n768 simulations, the numbers in brackets are the $\chi^2$ values where the power spectra are calculated using the rolling mean.}
\label{tab:chi2}
% z=5.4
\begin{tabular}{lcc}
\hline
name & min $\chi^2$ & UVB scaling \\% & UVB scaling \\
$z=5.4$ & $-2.3\le\log_{10}k({\rm s/km})\le-1.1$, d.o.f.$=6$ & at min $\chi^2$ \\
\hline
L25n512 RT-late & 6 & 1.0 \\
L25n512 RT-early & 5 & 1.3 \\
L25n512 RT-extended & 5 & 1.2 \\
L25n512 FR-z6.7 & 5 & 1.1 \\
L25n512 FR-z8.0 & 6 & 1.4 \\
L25n512 FR-z10 & 8 & 1.7 \\
L25n512 FR-z12 & 9 & 1.8 \\
L37.5n768 RT-early & 5 (5) & 1.3 \\
L37.5n768 RT-extended & 5 (4) & 1.2 \\
\hline
\end{tabular}
% z=5.0
\begin{tabular}{lccc|c}
\hline
name & min $\chi^2$ & UVB scaling & min $\chi^2$ & $P=\exp\left(-\Delta\chi^2/2\right)$ \\
$z=5.0$ & $k\le0.1$~s/km, d.o.f.$=12$ & at min $\chi^2$ & $k\le0.2$~s/km, d.o.f.$=15$ & \\
\hline
L25n512 RT-late & 11 & 1.6 & 24 & 1.6 \\
L25n512 RT-early & 13& 2.1 & 26 & 0.6  \\
L25n512 RT-extended & 10 & 1.9 & 22 & 2.7 \\
L25n512 FR-z6.7 & 12 & 1.8 & 27 & 1 \\
L25n512 FR-z8.0 & 15 & 2.1 & 29 & 0.2 \\
L25n512 FR-z10 & 21 & 2.3 & 37 & 0.01 \\
L25n512 FR-z12 & 26 & 2.4 & 43 & 0.0009 \\
L37.5n768 RT-early & 14 (13) & 1.9 & 27 (25) & 0.4 \\
L37.5n768 RT-extended & 12 (12) & 1.9 & 23 (21) & 1 \\
\hline
\end{tabular}
% z=4.6
\begin{tabular}{lccc|cc}
\hline
name & min $\chi^2$ & UVB scaling & min $\chi^2$ & $\chi^2_{\rm tot} = \sum_{z=4.6}^{5.4} \chi^2(z) $ & $P=\exp\left(-\Delta\chi^2_{\rm tot}/2\right)$ \\
$z=4.6$ or total & $k\le0.1$~s/km, d.o.f.$=12$ & at min $\chi^2$ & $k\le0.2$~s/km, d.o.f.$=15$ & d.o.f.$=30$ & \\
\hline
L25n512 FR-z6.7 & 6 & 1.5 & 32 & 23 & 1 \\
L25n512 FR-z8.0 & 9 & 1.7 & 36 & 30 & 0.03 \\
L25n512 FR-z10 & 17 & 1.8 & 46 & 46 & $10^{-5}$ \\
\hline
\end{tabular}
\end{table*}

We explore whether our patchy and uniform reionization models can be distinguished with the $z=5.4$ measurement of \citet{Viel13} and the $z=5.0$ (and $z=4.6$\footnote{We caution that heating due to HeII reionization in the FG09 UVB raises the IGM temperature by $\approx1700$~K at $z=4.6$. Our interpretations of the $z=4.6$ data are likely affected by the uncertainty associated with the HeII reionization history.}) one of \citet{Boera18}. We evaluate the minimum $\chi^2$ values for all simulated power spectra by varying $\langle F \rangle$ and using these measured power spectra as well as the bandpower covariances. For the \citet{Boera18} data, we use their resolution-corrected measurements and scaled the uncorrected covariance matrices\footnote{\url{https://arxiv.org/src/1809.06980v2/anc}} to get the resolution-corrected ones. $\langle F \rangle$ is changed in steps of $0.005$ for \citet{Viel13} and $0.01$ for \citet{Boera18}. These choices are such that $\chi^2$ differs by $\lesssim0.5$ from adjacent bins near the minimum.
For the \citet{Viel13} data, we use the $7$ measurements with $-2.3 \le \log_{10}k({\rm s/km}) \le -1.1$, matching the range used for their parameter analysis. For the \citet{Boera18} measurement, we report two sets of $\chi^2$ values, one computed using only the 13 data points with $k\le0.1$~s/km, and the other using all 16 data points with $k\le0.2$~s/km.  While our conclusions do not change significantly between the two sets of $\chi^2$ values, the latter set are high for 15 degrees of freedom, likely indicating some systematic (see below).  Table~\ref{tab:chi2} lists the minimum $\chi^2$ values.\footnote{At $z=5.4$, $5.0$, $4.6$, the values of $\langle F \rangle$ at minimum $\chi^2$ in different simulations range from $0.060-0.075$, $0.16-0.20$, and $0.21-0.25$, respectively.} 

There are a few technical aspects to address.  First, we find that generating mock spectra in different directions can change the $\chi^2$ values by up to 1, so we do not report numbers after the decimal point.  %In addition, the power in a redshift bin is averaged over a finite redshift with non-uniform coverage, but we only use a snapshot at the redshift midpoint of each redshift bin to calculate the power spectrum.  %Additionally, properly accounting for the rolling mean marginally improves our fits (see Table~\ref{tab:chi2}, although note that we find a larger effect than \citealt{Boera18} in Appendix~\ref{sec:rollingmean}). %Averaging the power in a $k$ bin, especially at $k>0.1$~s/km, might also help bring the simulated high-$k$ power into closer agreement with the observations. 
We find similar changes in $\chi^2$ can result from our approximation of the the $z=5.0$ power spectrum with the power spectrum of the snapshot at its mean redshift.\footnote{To find this, we assumed $2/3$ of $z=5$ power is from the $z=4.8$ snapshot and $1/3$ from $z=5.2$ to emulate the \citet{Boera18} redshift sampling (and with both snapshots assuming the same photoionization rate that is adjusted to produce the mean flux). We found differences in $\chi^2$ values at the 1-2 level from using a single $z=5.0$ snapshot, differences that do not change our results.} %Nor did we find any differences in the $\chi^2$ values by averaging the power in the $k$ bins using the $z=5.0$ snapshot. %However, these are simplistic estimates and likely miss out a lot of information obtained with the more sophisticated modelling in \citet{Boera18}.
 %While  effects may only be modest, they could add up to be significant at the level as the data are able to distinguish $5\%$ differences. 
Finally, for the two L37.5n768 simulations, the numbers in brackets are the $\chi^2$ values where the power spectra are calculated using the rolling mean instead of the global mean. We find that the rolling mean gives a smaller $\chi^2$ by $\Delta\chi^2=1-2$.

The Akaike information criterion specifies that the probability that one model is favored over the other is given by $\exp(-\Delta\chi^2/2)$, where $\Delta\chi^2$ is the difference of $\chi^2$ between two models. We will use the phraseology that a model is ``favored at $\sqrt{\Delta \chi^2}\sigma$''.
At $z=5.4$, all $\chi^2$ are within three of each other, indicating that no model is strongly preferred by this data set. At $z=5.0$ and $4.6$ the differences are more interesting. We first focus on the FR simulations. When evaluated with the $k\le0.1$~s/km data points, $\chi^2$ values in FR-z6.7 are smaller by $3$ than those in FR-z8.0, and by $\ge9$ than those in FR-z10 at these redshifts. This implies that in each redshift bin the FR-z8.0 and FR-z10 models would be disfavored compared to FR-z6.7 model by $1.7\sigma$ and $\ge3\sigma$, respectively.
Adding up the $\chi^2$ values of each model from the different redshift bins enlarges the $\chi^2$ differences, allowing FR-z6.7 to be favored over the FR-z8.0 models at $2.5\sigma$ level. The fifth column of the $z=4.6$ table illustrates this, which lists $\chi^2_{\rm tot} = \chi^2(z=5.4) + \chi^2(z=5.0) + \chi^2(z=4.6)$ for the three FR simulations. The last column shows $P=\exp\left(-\Delta\chi^2_{\rm tot}/2\right)$, where $\Delta\chi^2_{\rm tot}$ is the difference in $\chi^2_{\rm tot}$ of each simulation with respect to FR-z6.7. The probabilities that FR-z8.0 and FR-z10 are more favored by data from all three redshift bins than FR-z6.7 are thus $0.03$ and $10^{-5}$, respectively.

At $z=5.0$, similar $\chi^2$ differences are seen in the RT simulations as between the FR simulations with the same reionization midpoints. The differences in $\chi^2$ owe primarily to the redshift of reionization. Thus although we have not run our RT simulations to $z=4.6$, we suspect that, like the FR models, RT models with reionization midpoints $\gtrsim8$ are moderately ruled out by existing data.

\begin{figure}
\centering
\includegraphics[width=\columnwidth]{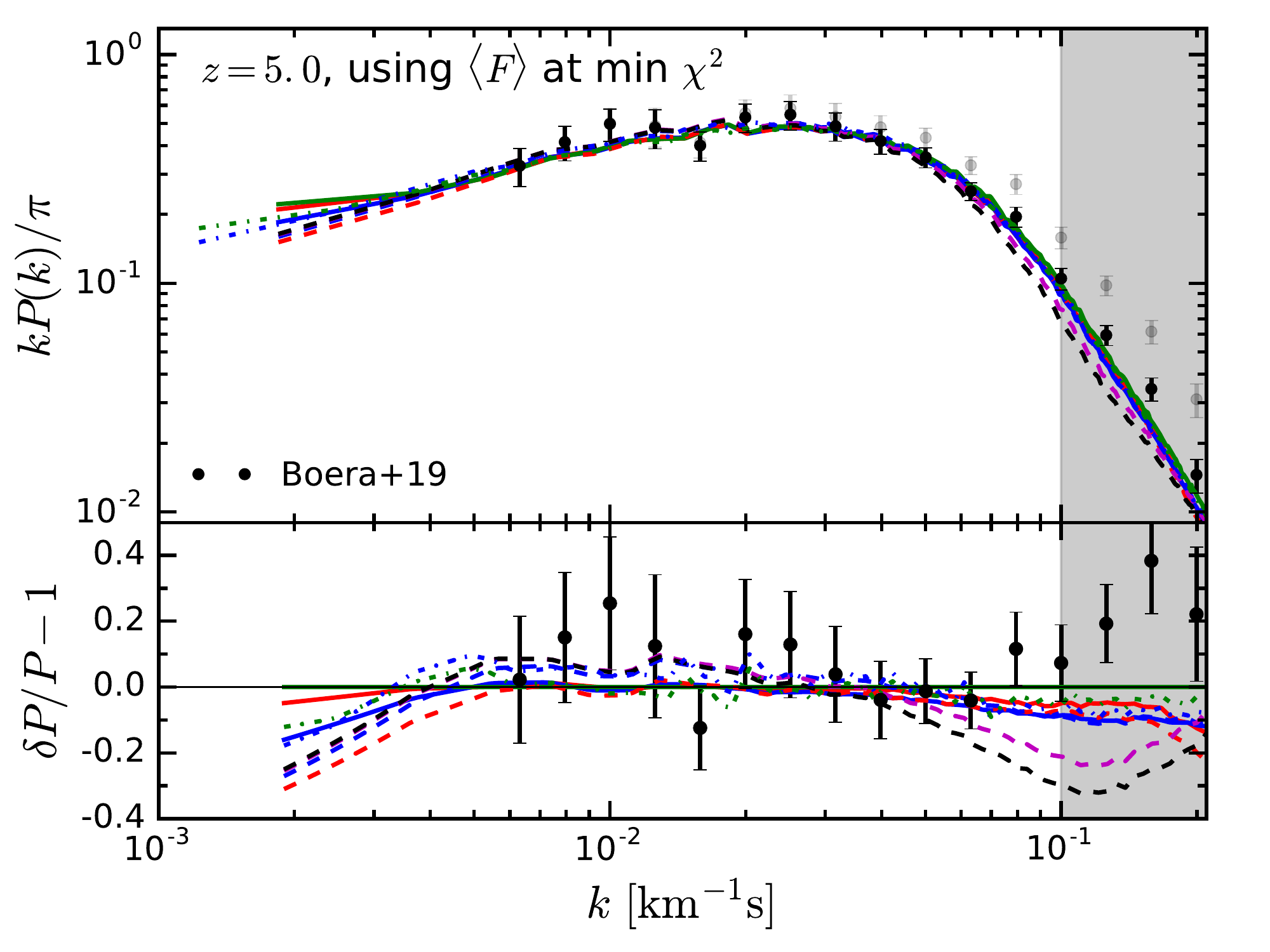}
\caption{$z=5.0$ power spectra in all simulations compared to the observational data of \protect\citet{Boera18} (black dots), each normalized to the mean flux at minimum $\chi^2$ (see Sec.~\ref{sec:data} for details). Colors and linestyles are the same as in Fig.~\ref{fig:powerspectrum}. Bottom panel shows the fractional differences in the power spectra compared to the one in L25n512 RT-extended, which has the smallest $\chi^2$. Shaded regions at $k>0.1$~s/km indicate the wavenumbers that are not fully used in our analysis. The larger gradient of power from $k\sim0.01$~s/km to $0.1$~s/km in models with higher reionization midpoints is the reason why these models are disfavored by the data.}
\label{fig:minchi2}
\end{figure}

Fig.~\ref{fig:minchi2} shows power spectra in all simulations at $z=5.0$, each normalized to the mean flux at minimum $\chi^2$. Colors and linestyles are the same as in Fig.~\ref{fig:powerspectrum}. The bottom panel shows the fractional differences in the power spectra, compared to the one in L25n512 RT-extended. This simulation has the smallest $\chi^2$ value at $z=5.0$. The shaded regions at $k>0.1$~s/km indicate the range of wavenumbers that are not fully used in our analysis. As the reionization midpoint gets higher, the power spectrum shows a larger gradient from $k\sim0.01$~s/km to $0.1$~s/km. This makes it difficult for earlier reionization models to match the data on all scales simultaneously. This illustrates the physical reason why models with reionization midpoints $z>8$ are less favored by the data than the later-ending reionization models.

Interestingly, our RT simulations provide a marginally better fit to the $z=5.0$ measurements than the FR simulations. The L25n512 RT-late and RT-early simulations have $\chi^2$ values for the case $k\le0.1$~s/km that are generally smaller by $1-2$ than FR-z6.7 and FR-z8.0. In addition, RT-extended is favored over RT-early at a similar level, although the L37.5n768 simulations produce $\chi^2$ values that are $1-2$ units higher than L25n512 due to slightly lower high-$k$ power. The last column of the $z=5.0$ table lists $P=\exp\left(-\Delta\chi^2/2\right)$, where $\Delta\chi^2$ is the $\chi^2$ difference of each simulation with FR-z6.7. While $\Delta\chi^2\sim1-2$ is not statistically significant, these differences suggest that more precise Ly$\alpha$ forest measurements could detect the signatures of a patchy reionization. Including data points at higher $k$ or modelling the redshift evolution of $\langle F \rangle$ with fewer parameters could also increase the significance level at which models can be distinguished.

When using all $16$ measurement points from \citet{Boera18}, each of our models has $\chi^2>22$. For $15$ degrees of freedom, the probabilities of getting $\chi^2 > 22$ is $0.1$.  While this could be a statistical fluke, this possibility is unlikely because the large increase in $\chi^2$ occurs from adding just three additional data points.  The correction we make to our simulations from resolution is largest at these high wavenumbers and, hence, is most uncertain there. A factor of $\sim1.5$ larger resolution correction than we applied is required to fit the measurements at $k\approx0.2$~s/km. While such a correction is somewhat larger than the we would expect at $z=5.0$ for these wavenumbers (see Fig.~\ref{fig:photoheating_convergence}), it is not implausibly large.  However, resolution effects are less severe at $z=4.6$, yet the $\chi^2$ values are even higher ($\ge32$) at $z=4.6$, suggesting that resolution may not be the full story.
% These highest wavenumbers are also most affected by metal contamination (as well as sharp cuts to excise metals). 
Because of these concerns, we focus on the $k\le0.1$~s/km $\chi^2$ values, although we note that the conclusions drawn from using all of $k\le0.2$~s/km data points are generally similar.

In addition to the $\chi^2$ values, Table~\ref{tab:chi2} lists the scaling factors of the FG09 UVB to get the minimum $\chi^2$ using the $k\le0.1$~s/km measurements (third column). The UVB scaling at minimum $\chi^2$ in models with reionization redshifts $\le8$ (the models that we find are most consistent with the measurements) give a sense for the allowed range of UVB intensities, although we caution that we do not have a full exploration of the parameter space. This complements the standard method for estimating the UVB from the mean flux for which the dominant uncertainty is the thermal history \citep[e.g.][]{Beck13}. %More importantly, the power spectrum puts constraints on the IGM thermal history, which is a major source of uncertainty in UVB measurements.
Our simulations bracket the UVB photoionization rate, $\Gamma$, to be $[3.0,4.6], [5.7,7.4], [5.8,6.7] \times10^{-13}$~s$^{-1}$ at $z=5.4, 5.0, 4.6$, respectively.  Our relatively small error bars on these quantities owes to the small differences in temperatures between our simulations after reionization, which in turn owes to the aysmptotic behavior of the post-reionization thermal history. With a softer ionizing background and eliminating HeII photoheating, we suspect that $2000$~K lower temperatures might be achievable compared to our simulations and, hence, $\approx 20\%$ smaller $\Gamma$ than our bounds. These values are broadly consistent with previous works. For instance, the measurements of \citet{Calv11} and \citet{Wyit11} using the quasar proximity region suggest that $\Gamma\approx4.5-10\times10^{-13}$~s$^{-1}$ at $z\approx5$, and \citet{DAlo18fluc} bracketed $\Gamma = 3.4-6.2\times10^{-13}$~s$^{-1}$. %$\log(\Gamma)=-12.15\pm0.16$ and $\Gamma = 0.47^{+0.3}_{-0.2}\times10^{-12}$ at $z\approx5$, respectively.
\citet{Beck13} estimated $\Gamma= 7.2-13.2\times10^{-13}$~s$^{-1}$ and $6.7-13.4\times10^{-13}$~s$^{-1}$ at $z=4.4$ and $4.75$, respectively.

\subsection{Intensity fluctuations}
\label{sec:UVB_fluc}

\begin{figure*}
\includegraphics[width=2\columnwidth]{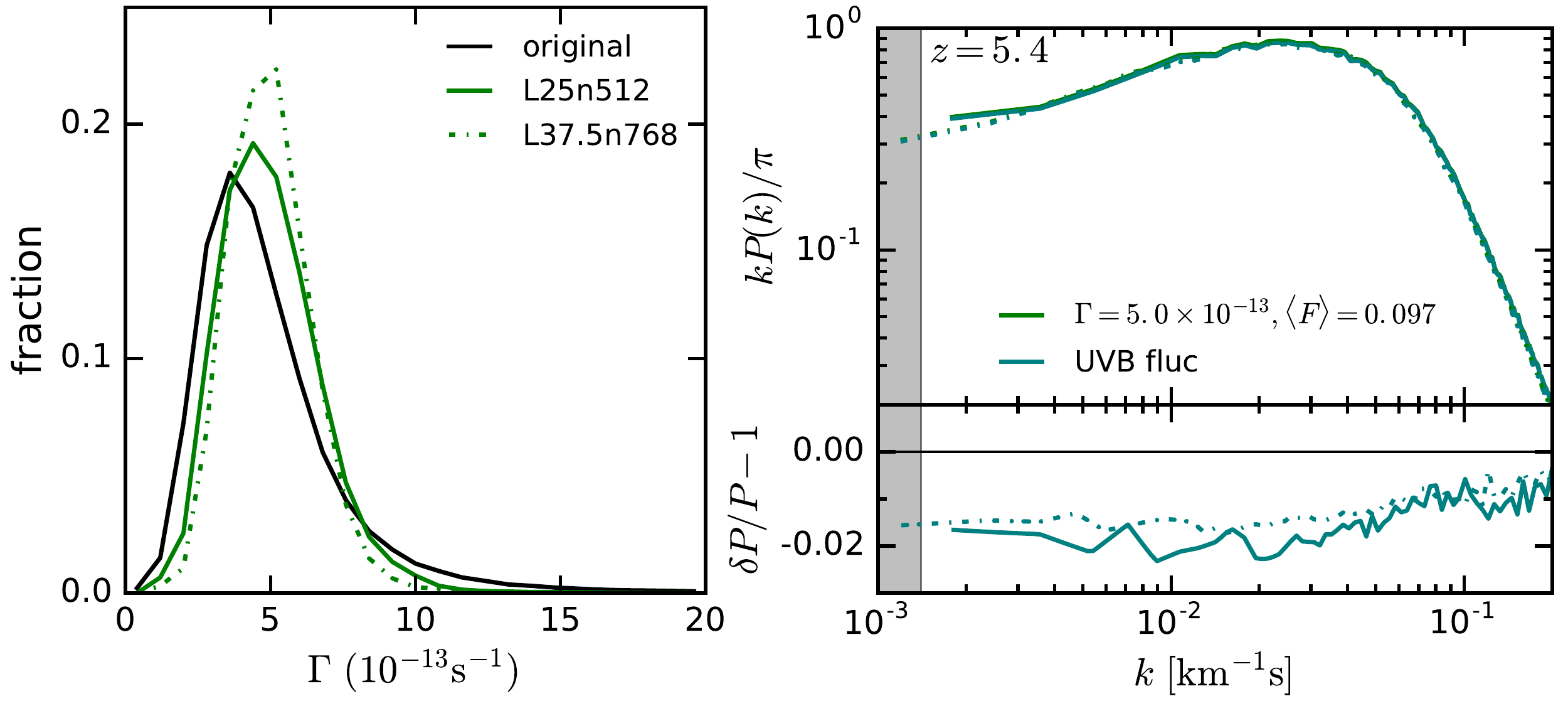}
\caption{An estimate of the effect of UV background fluctuations on the power spectrum. Left: black line shows the distribution of $\Gamma$ at $z=5.4$ in the ``short mean free path'' model of \protect\citet{DAlo18fluc}. Green solid and dot-dashed lines are the distributions after the original $\Gamma$ grid is convolved with top-hat kernels with volumes of $(25\, h^{-1}\ {\rm Mpc})^3$ (solid line) and $(37.5\, h^{-1}\ {\rm Mpc})^3$ (dot-dashed line), respectively. Right: the original (green lines) $z=5.4$ power spectra in L25n512 (solid) and L37.5n768 (dot-dashed) RT-extended simulations and those with UVB fluctuations (teal lines), assuming that the universe consists of patches with different local $\Gamma$ and that the global power spectrum is a weighted average of the local power spectrum (see text for a detailed description). The fractional differences shown in the bottom panel are calculated by the ratio of the power spectra with UVB fluctuations to those without. UVB fluctuations lower the amplitude of the power spectrum by $\lesssim2\%$, which is a much smaller effect than that of $T$ fluctuations.  While we expect our approximation of a uniform background on the size of our hydro simulation applies at most wavenumbers, at $k<0.0014$~s/km this approximation likely misses large-scale power contributed by the intensity fluctuations \protect\citep{DAlo18fluc}. This region is shown in gray.}
\label{fig:UVBfluc}
\end{figure*}

In addition to $T$ fluctuations, UV background fluctuations can also alter the shape of the power spectrum \citep[e.g.][]{DAlo18fluc, Onor18}. Because our simulation box sizes are smaller than the photon mean free path at $z\sim5$ \citep{Wors14}, even when doing full RT we would miss most of the contribution to intensity fluctuations because of our box size. Therefore, we present an estimate of the small-scale effects of UVB fluctuations on the Ly$\alpha$ forest power spectrum. In order to maximize the effects, we sample the photoionization rate $\Gamma$ from the ``short mean free path'' model of \citet{DAlo18fluc} and compute an average power spectrum. This model is computed in a $200\, h^{-1}$ Mpc simulation box on a $64^3$ grid. It has a mean $\Gamma$ of $5\times10^{-13}\ {\rm s^{-1}}$ and an average mean free path of $17 h^{-1}$ Mpc at $z=5.4$, which is a factor of 2 smaller than what the observations of \citet{Wors14} indicate. We convolve the original $\Gamma$ grid with top-hat kernels of the same volumes as the L25n512 and L37.5n768 simulations.
The left panel of Fig.~\ref{fig:UVBfluc} shows the original distribution function of $\Gamma$ at $z=5.4$ (black line), and those convolved with $(25\, h^{-1}\ {\rm Mpc})^3$ (solid line) and $(37.5\, h^{-1}\ {\rm Mpc})^3$ (dot-dashed line) top-hat kernels.

We apply the convolved $\Gamma$ grids to the L25n512 and L37.5n768 RT-extended simulations at $z=5.4$. We take 10 evenly-spaced values of $\Gamma$ from the distribution functions. For each $\Gamma$, the optical depth $\tau$ of each pixel is scaled by $\Gamma_{\rm FG09}/\Gamma$, where $\Gamma_{\rm FG09}$ is the photoionization rate given by the FG09 UVB. At $\Gamma=5\times10^{-13}$~s$^{-1}$, the RT-extended simulations have $\langle F \rangle = 0.097$. Assuming that the universe has $\langle F \rangle_{\rm global} = 0.097$ but is composed of patches with different local $\Gamma$, the power spectrum with UVB fluctuations is an average of the local power spectra, weighted by the distribution function of $\Gamma$. This integral over $\Gamma$ is performed using the midpoint rule with the 10 sampled points.
The right panel of Fig.~\ref{fig:UVBfluc} presents the power spectra with UVB fluctuations (teal lines), which almost overlap completely with the original ones (green lines). The bottom panel shows the fractional differences of the power spectra with UVB fluctuations to those without. The effect of UVB fluctuations is very small, reducing the amplitude of the power spectrum by $\lesssim2\%$. This is much smaller than the effects of $T$ fluctuations.

Our calculations ignore the correlations between the UVB and the density field, so is only relevant at high-$k$. Because the box sizes are smaller than the observed mean free path, the UVB is approximately homogeneous on the box size scale. Moreover, the similarity of the convolved and original distributions of $\Gamma$ shows that most of the variation is occurring on scales larger than our simulation box, justifying the approach taken here.
At $k<0.0014$~s/km, large-scale UVB fluctuations can also enhance large-scale power \citep{DAlo18fluc}. This low-$k$ region is colored in gray in Fig.~\ref{fig:UVBfluc}. However, the effects of large-scale UVB fluctuations can act against those of $T$ fluctuations and cancel out some large-scale excess power, because the UVB intensity is higher in denser regions where temperatures are lower due to earlier reionization \citep[e.g.][]{Onor18}. The inclusion of large-scale UVB fluctuations may therefore help reconcile the over production of low-$k$ power in our RT simulations. This is also more important at $z=5.4$, where large-scale opacity fluctuations are more likely driven by fluctuations in the UVB rather than temperature \citep{Beck18}. Nevertheless, it is possible that both $T$ fluctuations and UVB fluctuations play important roles in shaping the Ly$\alpha$ forest opacity fluctuations. As pointed out by \citet{Kulk19}, for a reionization that proceeds to $z<5.5$, the last neutral islands to be reionized attain the highest temperatures and thus exhibit the lowest opacity shortly after reionization. Local $T$ fluctuations hence wins over larger-scale UVB fluctuations in this case.

\section{Conclusions}
\label{sec:conclusions}
In this work we studied the imprints of post-reionization IGM temperature fluctuations on the shape of the $z\sim5$ Ly$\alpha$ forest flux power spectrum. This is the first time that this topic has been explored using fully coupled radiation-hydrodynamic simulations of reionization.  Our sources of ionizing radiation come from a state-of-the-art galaxy formation model. We simulated three different reionization histories that are consistent with current CMB measurements (a late-ending, early-ending, and extended model), which lead to different levels of temperature fluctuations. In conjunction, we ran a set of flash reionization simulations with tight temperature--density relations, which allowed us to both isolate effects and comment on importance of using RT to study reionization. Our primary conclusions are:

\begin{itemize}
\item
All of our simulations that span Planck's $\pm 1 \sigma$ range in electron scattering optical depth produce similar intermediate and small-scale power ($\lesssim20\%$ differences). This similarity owes to a surprisingly well-matched cancellation between thermal broadening and pressure smoothing effects for more physically motivated thermal histories.  Capturing this cancellation requires self consistent simulations like those presented here; post-processing simulations with the same ionization histories would likely result in larger differences.

\item
When the reionization midpoint is fixed, differences in the small-scale power generated by flash and patchy reionization models matched are less than $\sim20\%$. This result, which agrees with \citet{Onor18}, indicates that inhomogeneous heating in the patchy models does not substantially affect the $z\sim5$ Ly$\alpha$ forest power spectrum.

\item
We find that current measurements of the $z=5.4, 5.0, 4.6$ Ly$\alpha$ forest power spectrum constrain the reionization midpoint to be $z\le8$ at $2.5\; \sigma$ ($z\le 10$ at $5\; \sigma$). Interestingly, the observations also favor very modestly the patchy reionization models (especially our extended model) over the instantaneous models, suggesting that more precise measurements could detect the signatures of patchy reionization.

\item
The large-scale coherence of the temperature fluctuations has the strongest imprint on the Ly$\alpha$ forest power spectrum, bringing $20-60\%$ extra power at $k\sim0.002$~s/km and $z=5.0-5.4$ in our three RT models. This effect is most prominent in the late-ending and extended reionization simulations, where temperature fluctuations are the largest. This enhancement in power is qualitatively in agreement with previous works \citep{Cen09, DAlo18fluc, Onor18, MC19}. %However, the exact amount of excess power is smaller by $10-20\%$ in larger box simulations ($37.5\, h^{-1}$ Mpc) compared to ($25\, h^{-1}$ Mpc) simulations, which may indicate a lack of convegence.

\item
At small scales ($k\gtrsim0.1$~s/km), temperature fluctuations suppress power by $\lesssim10\%$ instead of raising it, because hotter regions with more transmission dominate the contribution to the global power spectrum. We illustrated this dependence with a simple model. This should result in an \emph{underestimate} in the warm/fuzzy dark matter mass (rather than an overestimate as had been hypothesized) when using simulations that do not include temperature fluctuations; this bias is small for current analyses.

\item
Our simulations with reionization midpoints $\le8$ allow us to constrain the background photoionization rate to be $\Gamma =[3.0,4.6], ~[5.7,7.4],~ [5.8,6.7] \times10^{-13}$~s$^{-1}$ at $z=5.4, 5.0, 4.6$, respectively. The small error bar arises from the lack of freedom in the thermal history $\Delta z\gtrsim 0.5$ after reionization.

\item
We investigated also the other source of large-scale opacity fluctuations, UV background fluctuations. We find that UV background fluctuations lower the amplitude of the power spectrum by $\lesssim2\%$ at $k>0.01$~s/km.  This is even much smaller than the impact of temperature fluctuations.  Our methodology does not allow us to address how UV background fluctuations enhance the power at lower wavenumbers.  
\end{itemize}

With the advent of more accurate data and better simulations, more information on reionization can be obtained from the power spectrum of the Ly$\alpha$ forest. Our simulations indicate that a factor of $\sim1.5$ reduction in the observational error bars would be able to distinguish most of our reionization models. Including measurements at $k>0.1$~s/km could also increase the significance level to differentiate models. In addition, measurements performed at lower $k$ should help distinguish patchy and uniform reionization scenarios.
Interestingly, our RT simulations overshoot the power at lowest wavenumbers that have been observed owing to large-scale temperature fluctuations. We were hesitant to focus on this discrepancy as we were unsure our simulations were of large enough volumes to be converged. It would be useful to run $\gtrsim100$ Mpc boxes to test convergence. Moreover, large boxes are able to simulate $T$ fluctuations and UVB fluctuations at the same time, potentially preventing an overshoot of the measured low-$k$ power owing to these two effects' cancellation.

A more careful future analysis should be done to resolve the discrepancy that we find between our simulated power spectra and the observational data at $k>0.1$~s/km and to further test our conclusions. Our work suggests that using the data at $k<0.1$~s/km is starting to be able to distinguish between interesting reionization models, but the shape of the power spectrum at $k=0.1-0.2$~s/km should add constraining power by helping break the degeneracy between thermal broadening and pressure smoothing effects \citep{Nasir16, Onor17ps, Boera18}. Our analysis can be improved by running higher resolution simulations as our simulations lose convergence at the highest wavenumbers (see Appendix~\ref{sec:reso_corr}).
In addition, we think a more detailed comparison between simulations and observations that mimics the methods used to reducing observational data in the simulations is motivated to confirm our results.

\section*{Acknowledgements}

We would like to thank Daniel Eisenstein, Josh Speagle, and Nick Gnedin for helpful conversations.  We are especially grateful to Vid Ir\v si\v c for discussions on how to analyze the \citet{Boera18} data set that used knowledge that had been gained from work in preparation. We also sincerely appreciate the comments from Jamie Bolton, George Becker, and Elisa Boera on the first draft of our paper. The simulations were performed on the Harvard computing cluster supported by the Faculty of Arts and Sciences, and the Comet supercomputer at the San Diego Supercomputing Center as part of XSEDE project TG-AST160069. M.M. acknowledges support from
NSF awards AST~1514734 and AST~1614439 and NASA
ATP award NNX17AH68G. RK acknowledges support from NASA through Einstein Postdoctoral Fellowship grant number PF7-180163 awarded by the Chandra X-ray Center, which is operated by the Smithsonian Astrophysical Observatory for NASA under contract NAS8-03060. FM acknowledges support from the program ``Rita Levi Montalcini'' of the Italian MIUR.

%%%%%%%%%%%%%%%%%%%%%%%%%%%%%%%%%%%%%%%%%%%%%%%%%%



%\bibliographystyle{mnras}


%%%%%%%%%%%%%%%%%%%%%%%%%%%%%%%%%%%%%%%%%%%%%%%%%%

%%%%%%%%%%%%%%%%% APPENDICES %%%%%%%%%%%%%%%%%%%%%

\appendix

\section{Numerical convergence of the gas temperature}
\label{sec:T_convergence}
We first discuss the numerical convergence of the IGM temperature with respect to the number of frequency bins and the gas cell size using a photoheating test problem.
We place an ionizing source in an initially neutral homogeneous IGM with hydrogen number density $6.5\times10^{-5}\ {\rm g/cm^3}$ and solve for $x_{\rm \hi}$ and $T$. The source has a UV spectral slope of $-1.5$ and emits $5\times10^{52}$ ionizing photons s$^{-1}$. The gas is represented by a Cartesian grid, where we vary the cell sizes with $5,10,20,40$ proper kpc. $10$ proper kpc is the typical gas cell size at $\bar{\rho}$ at $z\sim6$ in L25n512 and L37.5n768, while gas with $\rho/\bar{\rho}=0.1$ has sizes $\sim20$ proper kpc. $10$ proper kpc is also the typical I-front width found by \citet{DAlo18Ifront}. For choosing the frequency bins, we fix the $[24.6,54.4]$ eV bin, and divide the $[13.6,24.6]$ eV frequency range into $1,2,16$ bins evenly spaced in logarithmic space. The gas is evolved for 10 Myr. 

\begin{figure*}
\includegraphics[width=2\columnwidth]{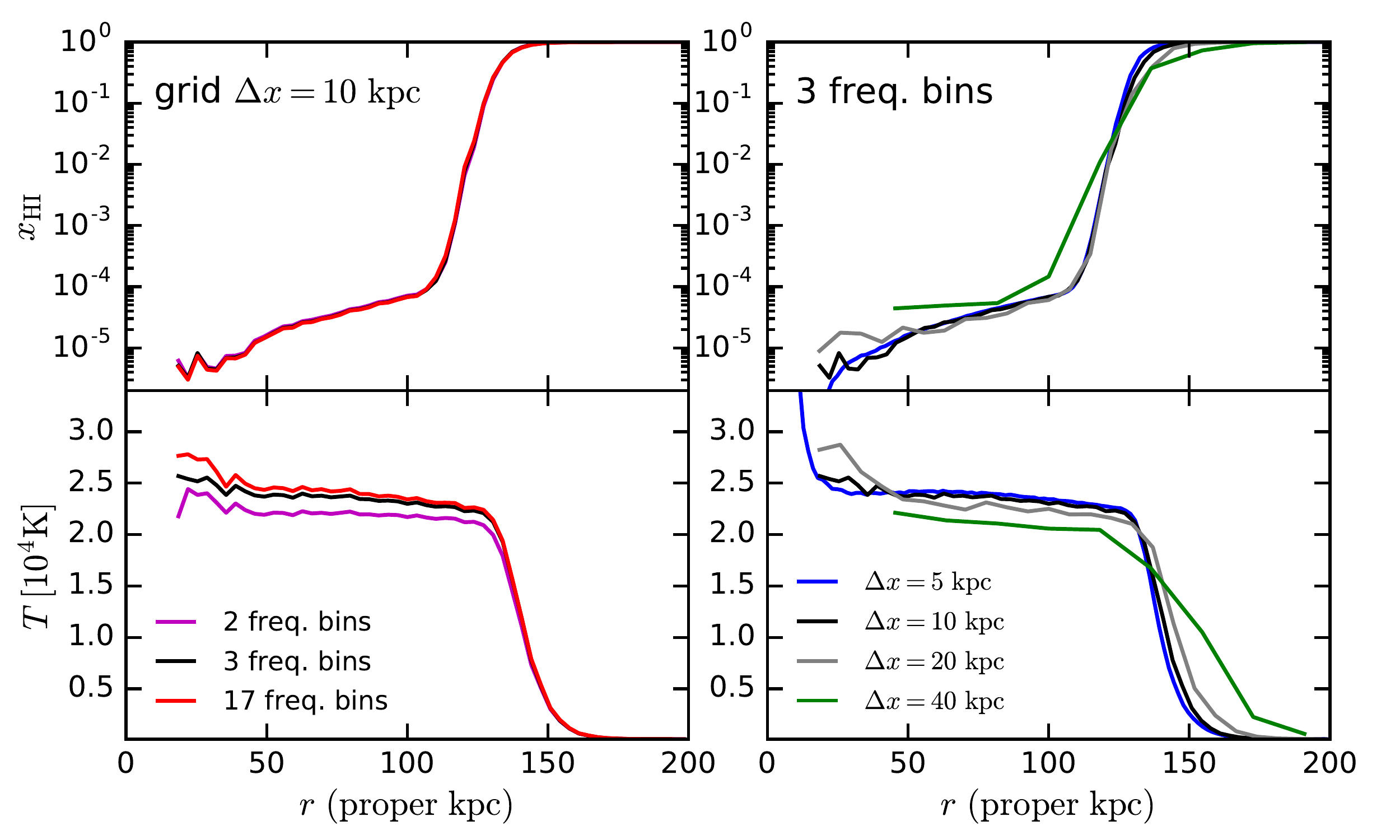}
\caption{Numerical convergence test with the number of frequency bins below $54.4$ eV and the size of the gas cells, for the case of a single ionizing source in an initially neutral homogeneous IGM. The source has a UV spectral slope of $-1.5$ and emits $5\times10^{52}$ ionizing photons s$^{-1}$. Top and bottom panels show the radial profiles of $x_{\rm \hi}$ and $T$ after evolving the gas for 10 Myr, respectively. Left: simulations have a grid size of $10$ proper kpc and $2,3,17$ frequency bins. Right: simulations use 3 frequency bins, with grid sizes of $5,10,20,40$ proper kpc. Overall, we find good numerical convergence with 3 frequency bins and cell sizes $<20$ proper kpc, indicating the IGM temperature in our simulations is well-converged.}
\label{fig:photoheating_convergence}
\end{figure*}

Fig.~\ref{fig:photoheating_convergence} presents the radial profiles of $x_{\rm \hi}$ (top panels) and $T$ (bottom panels) assuming the true speed of light for the calculation. The left panels show numerical convergence regarding the number of frequency bins, where the simulations adopt a grid cell size of $10$ proper kpc. The gas temperature is well-converged with 3 frequency bins of $[13.6,18.3], [18.3,24.6]$, and $[24.6,54.4]$ eV. The right panels show numerical convergence with respect to the grid cell size, where all simulations use 3 frequency bins. The radial profiles in the simulation with $40$ proper kpc cell size clearly deviate from the others, demonstrating the importance of correctly capturing the propagation of the I-front. Overall, the gas temperature is converged with cell sizes $<20$ proper kpc.

\begin{figure}
\includegraphics[width=\columnwidth]{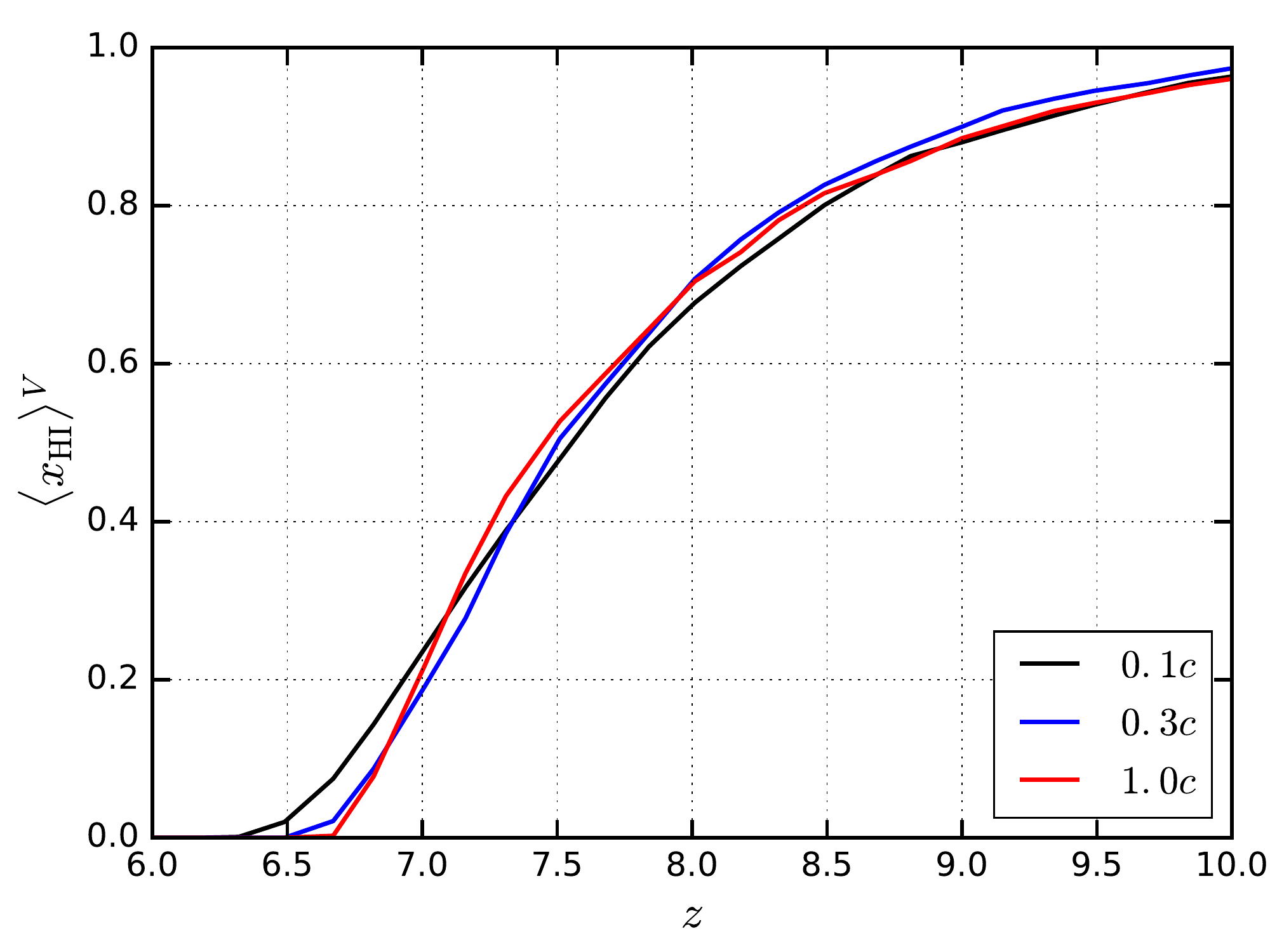}
\includegraphics[width=\columnwidth]{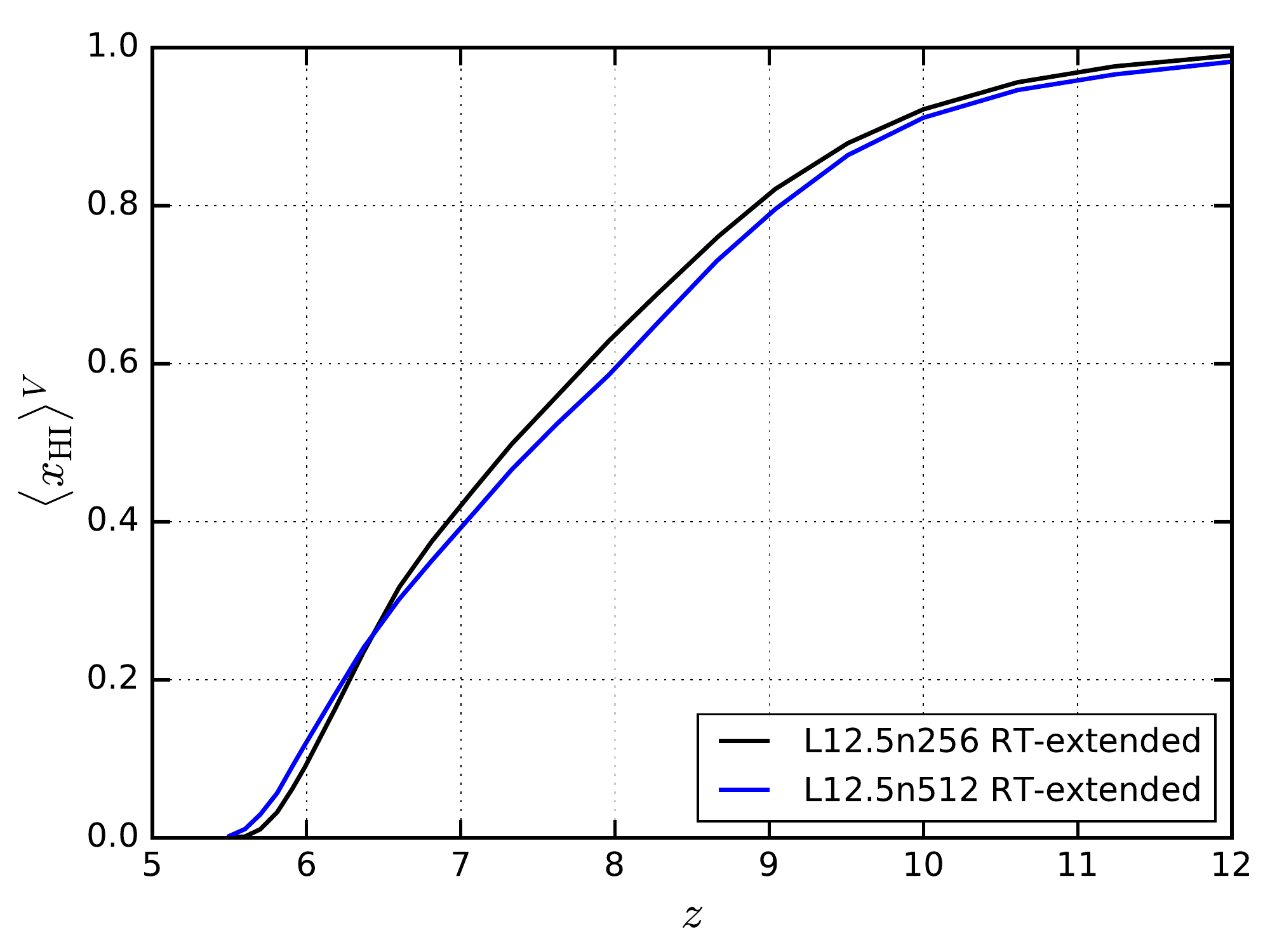}
\caption{Top: numerical convergence of the time evolution of the volume-averaged $x_{\rm \hi}$ with respect to the value of the reduced speed of light. Black, blue, and red lines represent L25n256 simulations run with $0.1c, 0.3c, 1.0c$, respectively. We use $0.1c$ for our simulations because of good convergence in the $x_{\rm \hi}$ evolution, therefore good convergence in the IGM temperature evolution.
Bottom: $\langle x_{\rm \hi} \rangle^V$ as a function of $z$ in L12.5n256 RT-extended and L12.5n512 RT-extended simulations. The escape fractions take the form of equation~\ref{eq:fesc}, except that the $f_{\rm esc}$ floor equals $0.35$. The reionization histories are similar in these two simulations.}
\label{fig:xHI_convergence}
\end{figure}

We now discuss the numerical convergence of the IGM temperature with respect to the choice of the reduced speed of light, using L25n256 simulations with $0.1c, 0.3c, 1.0c$ and $f_{\rm esc}=1$.
The top panel of Fig.~\ref{fig:xHI_convergence} shows the volume-averaged $x_{\rm \hi}$ as a function of redshift in the simulations. Since the $\langle x_{\rm \hi} \rangle^V$ evolution roughly reflects the average I-front speed, its good convergence indicates good convergence of the IGM temperature evolution. Moreover, \citet{DAlo18Ifront} found that the maximum I-front speed during reionization is likely $\sim0.1c$, suggesting that using a reduced speed of light of $0.1c$ is enough to correctly model the IGM temperature. Even though a reduced speed of light of $0.1c$ is not able to capture the high I-front speed tail near the end of reionization \citep{Depa19, Ocvi18}, the hottest regions cool down rapidly after being ionized. The post-reionization IGM temperature is therefore well-converged for $0.1c$. This is demonstrated by examining the IGM temperature--density diagrams at $z=6$ in the $0.1c$ and $1.0c$ simulations. We therefore adopt $0.1c$ for our simulations.

\section{Resolution correction of the power spectrum}
\label{sec:reso_corr}
We have run L12.5n256, L12.5n512, L12.5n640, and L12.5n768 simulations with the FG09 UV background to investigate the numerical convergence of the simulated power spectra under the mass resolution of L25n512. To check whether power spectra in the RT and FR simulations require the same amount of resolution correction, we also performed two RT-extended simulations with L12.5n256 and L12.5n512. The escape fractions take the form of equation~\ref{eq:fesc}, except that the $f_{\rm esc}$ floor is boosted to $0.35$ to ensure that reionization completes by $z=5.5$.
The bottom panel of Fig.~\ref{fig:xHI_convergence} shows the redshift evolution of $\langle x_{\rm \hi} \rangle^V$ in L12.5n256 (black line) and L12.5n512 (blue line) RT simulations. The reionization histories are similar in these two simulations, allowing the feasibility of exploring the dependence of the amount of resolution correction on the IGM thermal history.

\begin{figure*}
\includegraphics[width=2\columnwidth]{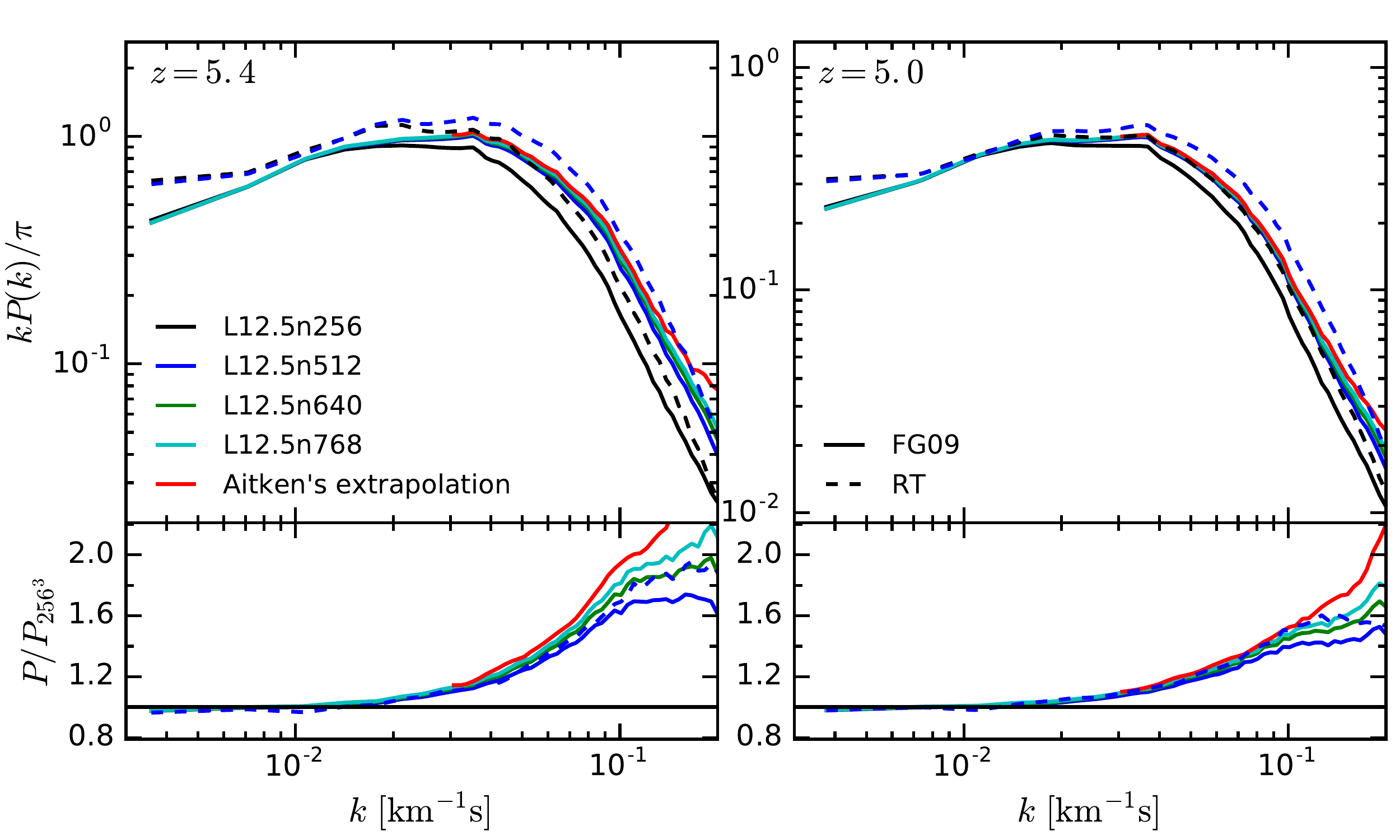}
\caption{Power spectra at $z=5.4$ (left) and $z=5.0$ in L12.5n256 (black solid lines), L12.5n512 (blue solid lines), L12.5n640 (green lines), and L12.5n768 (cyan lines) simulations that use the uniform FG09 UV background. Red lines show the extrapolated power spectra using L12.5n512, L12.5n640, L12.5n768, based on Aitken's delta-squared process. Dashed lines represent power spectra in L12.5n256 and L12.5n512 RT-extended simulations. All power spectra are normalized to the same mean fluxes: $\langle F \rangle = 0.080$ at $z=5.4$ and $\langle F \rangle = 0.184$ at $z=5.0$. The bottom panels show the ratios of the power spectra in the higher resolution simulations to those in L12.5n256. The resolution correction at $k>0.1$~s/km (where it is largest) is only relevant for analyzing the $z=5.0$ data.}
\label{fig:powerspectrum_resocorr}
\end{figure*}

Fig.~\ref{fig:powerspectrum_resocorr} illustrates the power spectra at $z=5.4$ (left) and $z=5.0$ (right) in different simulations. Solid and dashed lines represent FG09 and RT simulations, respectively. Black, blue, green, and cyan colors show power spectra in L12.5n256, L12.5n512, L12.5n640, and L12.5n768, respectively. The bottom panels show the ratio of each power spectrum with the one in L12.5n256 (L12.5nXXX FG09 divided by L12.5n256 FG09, L12.5n512 RT divided by L12.5n256 RT). Power spectra in the RT simulations seem to require $10-20\%$ more resolution correction at $k>0.1$~s/km than the FG09 simulations, but the overall agreement with the latter is relatively good. We therefore assume that power spectra in the L25n512 RT and FR simulations are converged at the same level.

Using Aitken's delta-squared process \citep{numerical_recipes}, we extrapolated the power spectra in L12.5n512, L12.5n640, L12.5n768 FG09 simulations to the ``true'' solution. This extrapolation method cancels out all leading order errors in the resolution corrections regardless of their scaling. The red lines in Fig.~\ref{fig:powerspectrum_resocorr} represent the resulting extrapolated power spectra at $k>0.03$~s/km. Power spectra in L12.5n768 are converged at $5-10\%$ level at $k=0.1$~s/km relative to the extrapolated power spectra. At $k<0.03$~s/km, the difference between the extrapolated power spectra and those in L12.5n768 is small.
Assuming the ``true'' power spectra consist of the L12.5n768 power spectra at lower $k$ and the extrapolated power spectra at high $k$ respectively, we drag down the observational data by the ratio of the L12.5n256 power spectra with the ``true'' solutions. We note that resolution correction at $k>0.1$~s/km does not apply to the $z=5.4$ data.

\section{Effects of varying the mean flux}
\label{sec:mf}

\begin{figure}
\includegraphics[width=\columnwidth]{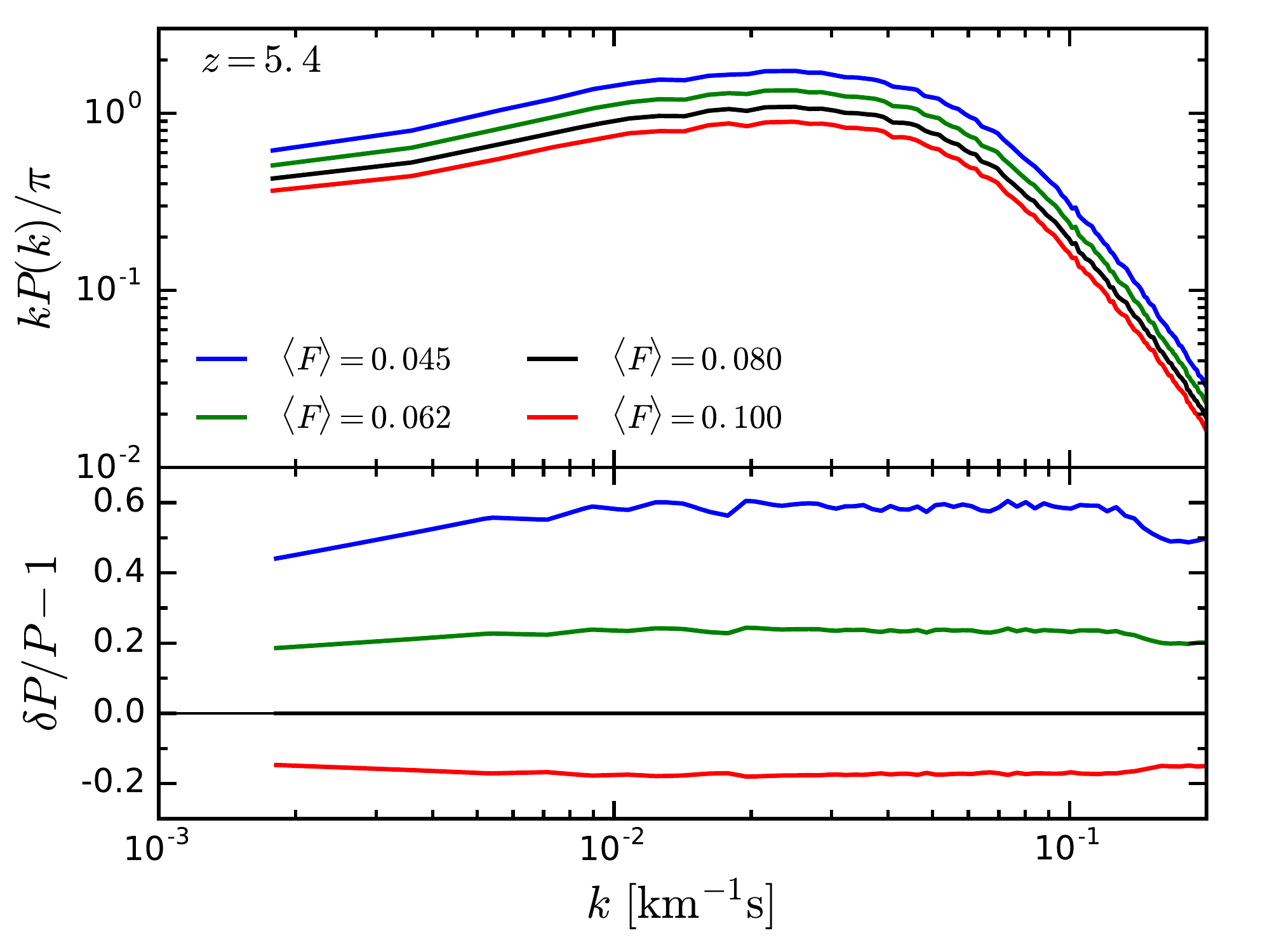}
\caption{The $z=5.4$ power spectra in L25n512 RT-late, normalized using different mean fluxes. Black line represents our default choice $\langle F \rangle = 0.08$, and the blue, green, and red lines show $\langle F \rangle = 0.046, 0.062, 0.010$ respectively. The bottom panel shows the fractional differences in the power spectra compared to the default one. Varying the mean flux changes the amplitude of the power spectrum but not its overall shape.}
\label{fig:powerspectrum_mf}
\end{figure}

In this section we consider the effects of varying the mean flux on the shape of the power spectrum. This illustrates whether our conclusions about the effects of temperature fluctuations are robust with respect to a change in the uncertain mean flux. To this end we calculate the power spectra in L25n512 RT-late with four different values of $\langle F \rangle$: $0.046, 0.062, 0.080, 0.010$. $0.046$ and $0.062$ are the values of $\langle F \rangle$ used in \citet{Viel13} and \citet{Keat18} respectively. $\langle F \rangle = 0.080$ is our default choice, and $\langle F \rangle = 0.010$ is about two sigma above the measured value in \citet{Bosm18}. Fig.~\ref{fig:powerspectrum_mf} compares these power spectra, and shows the fractional differences with respect to the default one in the bottom panel. Changing the mean flux mostly shifts the amplitude of the power spectrum, but not its shape. This is consistent with \citet{Boera18}. Thus the effects of the uncertain mean flux is not degenerate with the effects of temperature fluctuations.

\section{Effects of using the rolling mean flux}
\label{sec:rollingmean}

Here we present an estimate on the differences of using the rolling mean flux to calculate the power spectrum from using the global mean flux. Since our $25\, h^{-1}$Mpc boxes are too small compared to the $40\, h^{-1}$Mpc boxcar window in \citet{Boera18}, we use the L37.5n768 RT-extended simulation for this estimate. To replicate the effect of a rolling mean, we define the overdensity in flux using the mean flux of each sightline, rather than the mean flux in the entire box. This would emphasize contributions of sightlines with lower mean fluxes to the power spectrum, potentially affecting our analysis in Sec.~\ref{sec:thermalbroadening} (since we claimed that hotter regions dominate the contribution to the small-scale power). We perform this calculation using the original simulation outputs, and also with the gas cells put onto their mean $T-\rho$ relations. The purpose of the latter is to mimic an L37.5n768 FR simulation. An FR simulation should have an intrinsic distribution of mean fluxes of the sightlines due to density fluctuations, while $T$ fluctuations widen this distribution. The RT-extended simulation should give us an upper limit on the effect of the rolling mean, since this simulation has the largest $T$ fluctuations. For using both the global mean and the rolling mean, we rescale the optical depths of the pixels to obtain a desired global mean flux of $\langle F \rangle = 0.08$ at $z=5.4$ and $\langle F \rangle = 0.184$ at $z=5.0$.

\begin{figure*}
\includegraphics[width=2\columnwidth]{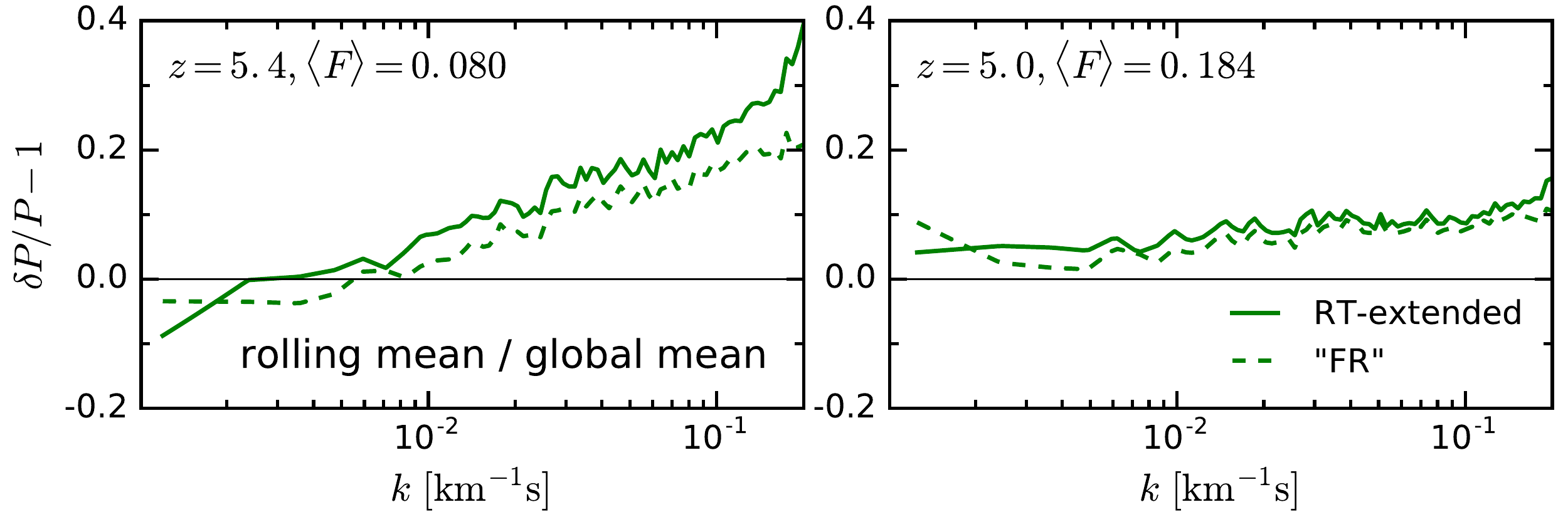}
\caption{Fractional differences of the power spectra in L37.5n768 RT-extended using the rolling mean versus using the global mean. Left and right panels show results at $z=5.4$ and $z=5.0$ respectively. In both methods of calculating the power spectrum, the optical depths of the pixels are scaled to obtain the global mean fluxes given on top of each panel. Solid lines use the power spectra calculated with the original outputs of the simulations, while the dashed lines use the power spectra where $T$ fluctuations are artificially removed by placing gas cells onto the mean temperature--density relations. The latter mimics a L37.5n768 FR simulation. Temperature fluctuations enlarge the variations in the mean fluxes of the sightlines, so using the rolling mean raises the small-scale power at $z=5.4$ compared to using the global mean.}
\label{fig:rollingmean}
\end{figure*}

Fig.~\ref{fig:rollingmean} shows the fractional differences in the power spectra using the rolling mean compared to those using the global mean. Solid and dashed lines represent results calculated with the original simulation outputs, and with the gas cells put onto the mean $T-\rho$ relations (denoted as ``FR'' in this section), respectively. Interestingly, at $z=5.0$ (right panel), the ``FR'' results indicate that using the rolling mean shifts the amplitude of the power spectrum up by $\sim10\%$ compared to using the global mean, but does not change its shape much. At $z=5.4$, the rolling mean raises the power at $k>0.1$~s/km by $\sim20\%$ in ``FR''. This is inconsistent with the findings of \citet{Boera18} that using the $40\, h^{-1}$Mpc boxcar window recovers the power on all scales as using the global mean. This is likely caused by the simplistic approach we adopt to calculate the rolling mean power spectrum. For instance, the power in a redshift bin is averaged over the bin with non-uniform coverage in \citet{Boera18}. They considered the impact of a rolling mean by stitching snapshots of their a $10\, h^{-1}$Mpc simulations, mimicking exactly how the rolling mean was applied to their observations, but we only use a snapshot at the redshift midpoint of each redshift bin to calculate the power spectrum. The differences may also owe to \citet{Boera18} not capturing $>10\, h^{-1}$Mpc correlations. We defer more detailed analysis towards understanding these issues to future work, but only focus on the differences between the RT power spectra and the ``FR'' ones. At $z=5.4$, using the rolling mean in RT increases the small-scale power by $\sim10-20\%$ more than in ``FR'', implying that temperatures fluctuations enlarge the differences in the mean fluxes of the sightlines. This also demonstrates that the small-scale power is determined by sightlines with lower mean fluxes when using the rolling mean. At $z=5.0$, the rolling mean changes the power spectra in RT and ``FR'' by almost the same amount. Hence our analysis regarding the $z=5.0$ results should be more robust.

\begin{figure*}
\includegraphics[width=2\columnwidth]{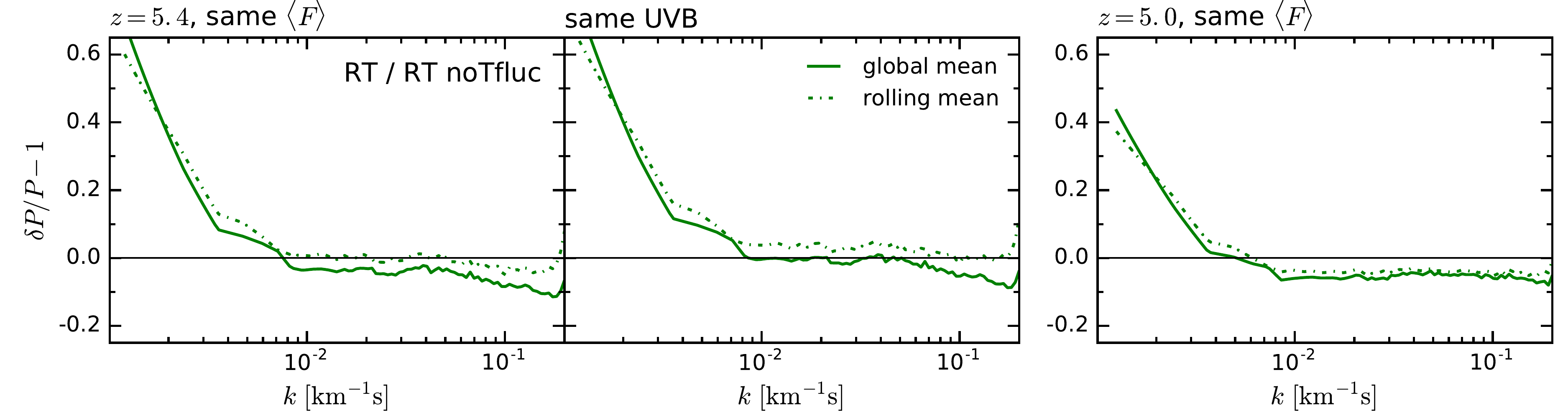}
\caption{Similar to the top row of Fig.~\ref{fig:residual}: power spectra in the L37.5n768 RT-extended simulations compared to those where $T$ fluctuations are artificially removed by placing gas cells onto their mean temperature--density relations. Solid and dot-dashed lines show results using the global mean and the rolling mean, respectively. Using the rolling mean mainly affects our interpretations on the effects of $T$ fluctuations in Sec.~\ref{sec:thermalbroadening} at $z=5.4$.}
\label{fig:rollingmean_noTfluc}
\end{figure*}

Fig.~\ref{fig:rollingmean_noTfluc} is a replicate of the top panels of Fig.~\ref{fig:residual}, where we show the fractional differences of the power spectra in RT compared to those in ``FR'' (where $T$ fluctuations are artificially removed). Solid and dot-dashed lines represent results using the global mean and the rolling mean, respectively. At $z=5.4$, using the rolling mean almost eliminate the suppression of the small-scale power by temperature fluctuations. However, due to the differences of the simplistic approach adopted here from that of \citet{Boera18}, we are also likely overestimating the effects of the rolling mean. Results at $z=5.0$ are not changed by using the rolling mean, but the suppression of the small-scale power by $T$ fluctuations is only $\sim5\%$. Our conclusions regarding the role of $T$ fluctuations are therefore affected by the method of calculating the power spectrum.

%%%%%%%%%%%%%%%%%%%%%%%%%%%%%%%%%%%%%%%%%%%%%%%%%%

% Don't change these lines
\bsp	% typesetting comment
\label{lastpage}
\end{document}